\documentclass[traditabstract]{aa}
\usepackage{amsmath}
\usepackage{graphicx}
\usepackage{natbib}
\usepackage{txfonts}
\usepackage{indentfirst}

\newcommand{\eq}[1]{\begin{equation}  #1 \end{equation}}

\newcommand{\eqa}[1]{\begin{eqnarray}   #1 \end{eqnarray}}

\newcommand{\br}[1]{\left( #1 \right)}
\newcommand{\bc}[1]{\left\{ #1 \right\}}

\newcommand{\ba}[1]{\left\langle #1 \right\rangle}

\newcommand{\vek}[1]{\mbox{\boldmath $#1$}}

\begin{document}

\title{Controlling intrinsic-shear alignment in three-point weak lensing statistics}

\author{X. Shi \inst{1,2}\and B. Joachimi \inst{1} \and P. Schneider \inst{1}}

\offprints{X. Shi,\\
    \email{xun@astro.uni-bonn.de}
}

\institute{Argelander-Institut f\"ur Astronomie (AIfA), Universit\"at Bonn, Auf dem H\"ugel 71, 53121 Bonn, Germany
\and International Max Planck Research School (IMPRS) for Astronomy and Astrophysics at the Universities of Bonn and Cologne}

\date{Received 3 February 2010 / Accepted 5 July 2010}

\abstract{
Three-point weak lensing statistics provide cosmic information that complements two-point statistics.  However, both statistics suffer from intrinsic-shear alignment, which is one of their limiting systematics. The nulling technique is a model-independent method developed to eliminate intrinsic-shear alignment at the two-point level. In this paper we demonstrate that the nulling technique can also be naturally generalized to the three-point level, thereby controlling the corresponding GGI systematics. 

We show that under the assumption of exact redshift information the intrinsic-shear alignment contamination can be completely eliminated. To show how well the nulling technique performs on data with limited redshift information, we apply the nulling technique to three-point weak lensing statistics from a fictitious survey analogous to a typical future deep imaging survey, in which the three-point intrinsic-shear alignment systematics is generated from a power-law toy model.

Using 10 redshift bins, the nulling technique leads to a factor of 10 suppression of the GGI/GGG ratio and reduces the bias on cosmological parameters to less than the original statistical error. More detailed redshift information allowing for finer redshift bins leads to better bias reduction performance. The information loss during the nulling procedure doubles the statistical error on cosmological parameters. A comparison of the nulling technique with an unconditioned compression of the data suggests that part of the information loss can be retained by considering higher order nulling weights during the nulling procedure. A combined analysis of two- and three-point statistics confirms that the information contained in them is of comparable size and is complementary, both before and after nulling.
}


\maketitle

\section{Introduction}
\label{sec:introduction}

Weak gravitational lensing refers to the mild distortion of the light from distant sources by the large-scale matter inhomogeneity between the source and the observer. One observable effect of weak gravitational lensing is the coherent shape distortion of the light sources, known as cosmic shear. Cosmic shear is sensitive to all cosmological parameters which have influence on the density perturbations and/or the geometry of the universe, including those concerning properties of dark energy, which have been a key concern after the discoveries made by observations of supernovae, the cosmic microwave background, and the large-scale structure \citep[for a review see e.g.][]{munshi08}.

Since its first detection in 2000 \citep{bacon00, kaiser00, vwaer00, wittman00}, cosmic shear has been developed into a competitive cosmological probe. Its constraining power on cosmological parameters is now comparable to other probes \citep[e.g.][]{spergel07,fu08}. 

With forthcoming large field multicolor imaging surveys (e.g.  DES\footnote{\texttt{http://www.darkenergysurvey.org/}}, KIDS\footnote{\texttt{http://www.astro-wise.org/projects/KIDS/}}, EUCLID\footnote{\texttt{http://sci.esa.int/science-e/www/area/\\index.cfm?fareaid=102}}, etc), photometric redshift and shape information of a huge number of galaxies will be available, rendering cosmic shear even greater statistical power. In particular, cosmic shear is considered to be one of the most promising dark energy probes \citep{TaskForce06,ESO-ESA06} when the results of these surveys become available.

Such constraining power will be further enhanced by the use of higher-order statistics. Second-order statistics measure only the Gaussian signature of a random field. Even if the primordial cosmic density field is Gaussian, non-Gaussianity will be generated due to the nonlinear nature of gravitational clustering. Such non-Gaussianity in the cosmic density field will then show up via its lensing effect, leading to non-Gaussian signals in the cosmic shear field. A common way of measuring non-Gaussianity is to use higher-order statistics. In cosmic shear studies, several authors have shown that the lowest order of them, i.e. the third-order statistics, already provide a valuable probe for cosmological parameter estimates; in particular it can break the near degeneracy between the density parameter $\Omega_{\rm m}$ and the power spectrum normalization $\sigma_{\rm 8}$ \citep{ber97,jain97,vwaer99,hui99}. A more recent study by \citet{TJ04} (TJ04 afterwards) showed that including third-order statistics can improve parameter constraints significantly, typically by a factor of three.

But the ultimate performance of these future surveys still largely depends on the how well the systematic errors can be controlled \citep[e.g.][]{hut06}. In this paper we focus on a particularly worrisome systematic error in cosmic shear studies: the intrinsic-shear alignment, and demonstrate a way to control it for shear three-point statistics.

In the weak lensing limit the observed ellipticity of a galaxy $\epsilon_{\rm obs}$ can be written as the sum of the intrinsic ellipticity $\epsilon_{\rm I}$ of the galaxy, and the shear $\gamma$ which is caused by gravitational lensing of the foreground matter distribution. Here $\epsilon_{\rm obs}$, $\epsilon_{\rm I}$ and $\gamma$ are complex quantities. Intrinsic-shear alignment is defined in two-point cosmic shear statistics as the correlation between the intrinsic ellipticity of one galaxy and the shear of another galaxy (the GI term, \citealp{hirata04}). Three-point statistics $\ba{\epsilon_{\rm obs}^i\epsilon_{\rm obs}^j\epsilon_{\rm obs}^k}$, a correlator of ellipticities of three galaxy images $i$, $j$ and $k$, can also be expanded into lensing (GGG), intrinsic-shear (GGI and GII), and intrinsic (III) terms:

\eq{
\label{eq:split}
\ba{\epsilon_{\rm obs}^i  \epsilon_{\rm obs}^j \epsilon_{\rm obs}^k}  = \rm{GGG} + \rm{GGI} + \rm{GII} + \rm{III} \;,\mbox{with} 
}

\eq{
\textrm{GGG} = \ba{\gamma^i \gamma^j \gamma^k} \;,
}

\eq{
\textrm{GGI} = \ba{\epsilon^i_{\textrm I} \gamma^j \gamma^k} + \ba{\epsilon^j_{\textrm I} \gamma^k \gamma^i} +\ba{\epsilon^k_{\textrm I} \gamma^i \gamma^j} \;,
}

\eq{
\textrm{GII} = \ba{\epsilon^i_{\textrm I} \epsilon^j_{\textrm I} \gamma^k} + \ba{\epsilon^j_{\textrm I} \epsilon^k_{\textrm I} \gamma^i} +\ba{\epsilon^k_{\textrm I} \epsilon^i_{\textrm I} \gamma^j} \;,
}

\eq{
\textrm{III} = \ba{\epsilon^i_{\textrm I}  \epsilon^j_{\textrm I} \epsilon^k_{\textrm I}} \;.
}

Physically, if one assumes that galaxies are randomly oriented on the sky, only the desired GGG term remains on the right-hand side of (\ref{eq:split}). However, when these galaxies are subject to the tidal gravitational force of the same matter structure (e.g. they formed under the influence of the same massive dark matter halo), their shapes can intrinsically align and become correlated, giving rise to a nonvanishing III term. Furthermore, GGI and GII terms can be generated when a matter structure tidally influences close-by galaxies and at the same time contributes to the shear signal of background objects, leading to correlations among them. 

In two-point statistics, the corresponding intrinsic (II) and intrinsic-shear (GI) terms have been subject to detailed studies both theoretically (e.g. \citet{catelan01, croft00, heavens00, hui02, mackey02, jing02, hirata04, heymans06, bridle07, schneider09}) 
and observationally \citep{brown02, heymans04, mandel06, mandel09, hirata07, fu08, brainerd09, okumura09a, okumura09b}. Although the results of these studies show large variations, most of them are consistent with a 10$\,\%$ contamination by both II and GI correlations for future surveys with photometric redshift information. Especially, neglecting these correlations can bias the dark energy equation of state parameter $w_0$ by as much as $50\,\%$ \citep{bridle07b} for a ``shallow'' survey described in \citet{amara07}. For three-point shear statistics, there have been few measurements up to now \citep{ber02,pen03,jarvis04}. However the potential systematics level in these studies is found to be high. A recent numerical study by \citet{sembo08} showed that intrinsic alignments affect three-point weak lensing statistics more strongly than at the two-point level for a given survey depth. In particular, neglecting GGI and GII systematics would lead to an underestimation of the GGG signal by ~$5-10\,\%$ for a moderately deep survey like the CFHTLS Wide. Therefore, to match the statistical power expected for cosmic shear in the future surveys, it is essential to control these systematics. 

The intrinsic alignment, II\,(III) in the two-\,(three-) point case, is relatively straightforward to eliminate, since it requires that the galaxies in consideration are physically close to each other, i.e. have very similar redshifts and angular positions \citep{king02,king03,HH03,TW04}. The control of intrinsic-shear systematics, GI for the two-point case and GGI in the three-point case (GII also requires that two of the three galaxies are physically close and thus can be eliminated in the same way as II and III), turns out to be a much greater challenge. However, as already pointed out by HS04, the characteristic dependence on galaxy redshifts is a valuable piece of information that helps to control the intrinsic-shear alignments.

Several methods for this have already been constructed in the context of two-point statistics. They can be roughly classified into three categories: modeling \citep{king05,bridle07b}, nulling (\citealp{JS08b}, JS08 hereafter; \citealp{joachimi09a}) and self-calibration \citep{zhang08, joachimi09c}. Modeling separates cosmic shear from the intrinsic-shear alignment effect by constructing template functions for the latter. It suffers from uncertainties of the model due to the lack of knowledge of the angular scale and redshift dependence of the intrinsic-shear signal. The nulling technique employs the characteristic redshift dependence of the intrinsic-shear signal to ``null it out''. It is a purely geometrical method and is model-independent, but suffers from a significant information loss. Self-calibration intends to solve the problem of information loss by using additional information from the galaxy distribution to ``calibrate'' the signal. The original form of self-calibration, proposed by \citet{zhang08}, is model-independent but strong assumptions have been made. \citet{joachimi09c} then develop it into a modeling method, by treating intrinsic alignments and galaxy biasing as free functions of scale and redshift.  

All these methods have the potential of being generalized to three-point statistics. In this paper we focus on the nulling technique, and establish it as a method to reduce the three-point intrinsic-shear alignments GGI and GII. Since GII can be removed by discarding close pairs of galaxies as in the case of II controlling \citep[e.g.][]{HH03}, we focus on the control of GGI systematics.

As known from the case of two-point statistics, the nulling technique introduces significant information loss while (in principle) completely removing the intrinsic-shear alignment from the signal. In this work we compare the nulling technique to an unconditioned linear compression of the data, distinguish different sources of such information loss, and discuss the possible ways of reducing it. We also study the combined constraints on cosmological parameters with both two- and three-point cosmic shear statistics.

In Sect.$\,$2 we demonstrate why and how the nulling technique can be applied to three-point lensing statistics. We then apply the nulling technique to the modeled lensing bispectrum which we contaminate by intrinsic-shear alignment. The modeling details are described in Sect.$\,$3. The method of nulling weights construction and the corresponding results are shown in Sect.$\,$4, while the results concerning the constraints on cosmological parameters are presented in Sect.$\,$5. We conclude in Sect.$\,$6.

We will work in the context of a spatially flat CDM cosmology with a variable dark energy whose equation of state $w$ is parameterized as $w = w_0 + w_a (1-a)$, with $a$ the cosmic scale factor. 
The adopted fiducial values for cosmological parameters are 
$\Omega_{\rm m}=0.3$, 
$\Omega_{\rm b}=0.045$,
$\Omega_{\rm de}=0.7$,
$w_0=-0.95$,  
$w_a=0.0$,
$h=0.7$,
$n_s=1.0$, and
$\sigma_8=0.8$. Here, $\Omega_{\rm m}$, $\Omega_{\rm b}$ and $\Omega_{\rm de}$ are the density parameters of the matter (including cold dark matter and baryons), baryons and the dark energy at present time, $n_s$ is the spectral index of the primordial power spectrum of scalar perturbations, $h$ is the dimensionless Hubble parameter defined by $H_0=100\, h\, {\textrm{km/s/Mpc}}$, and $\sigma_8$ is the rms mass fluctuation in spheres of radius $8h^{-1}$Mpc.

\section{The nulling technique applied to three-point shear tomography}
\label{sec:theory}

\subsection{Principle of the nulling technique}
\label{sec:nulling}

The shear on the image of a distant galaxy is a result of gravitational distortion of light caused by the inhomogeneous three-dimensional matter distribution in the foreground of that galaxy. For notational simplicity, we will use the dimensionless surface mass density (the convergence) $\kappa$ instead of the shear ${\gamma}$ as a measure for the lensing signal throughout the paper, although in reality the signal is based on the measurement of the shear. This will not affect our results since $\kappa$ and ${\gamma}$ are linearly related on each redshift plane while our method is dealing with the redshift dependence of them (the same reason justifies the turning to the Fourier domain in the next subsection). 

When one measures the shear ${\gamma}$, the direct observable is the galaxy ellipticity $\epsilon_{\rm obs}=\epsilon_{\rm I}+{\gamma}$. The shear ${\gamma}$ is a signal caused by gravitational distortion which is a deterministic process, where the intrinsic ellipticity $\epsilon_{\rm I}$ can be further written as the sum of a deterministic part $\epsilon_{\rm I}^{\rm det}$ which is caused by intrinsic alignment, and a stochastic part $\epsilon_{\rm I}^{\rm ran}$ which does not correlate with any other quantity. There is no correlation between $\epsilon_{\rm I}^{\rm ran}$ of different galaxies either.

We define $\kappa_{\rm obs}$ and $\kappa_{\rm I}$ which are the correspondences of $\epsilon_{\rm I}^{\rm det}+{\gamma}$ and $\epsilon_{\rm I}^{\rm det}$. We remove the stochastic part since $\kappa$ is deterministic. Note that $\kappa_{\rm obs}$ and $\kappa_{\rm I}$ are analogs of the dimensionless surface mass density $\kappa$ but do not have any direct physical meaning as $\kappa$ does. They are complex quantities in general and can lead to a B-mode signal. To better distinguish the real measurable $\kappa$ from them, we denote it as $\kappa_{\rm G}$ in the rest of the paper since it is the physical quantity which is related to the gravitational lensing signal. Keeping the dominating linear term, the convergence $\kappa_{\rm G}$ can be written as \citep[details see e.g.][]{s06}:

\eq{
\label{eq:d2k}
\kappa_{\rm G}(\vek{\theta},\chi_{\rm s}) = \frac{3 \, \Omega_{\rm m} H_0^2 }{2 c^2} \int_0^{\chi_{\rm s}} d\chi\, \frac{\chi (\chi_{\rm s}-\chi)}{\chi_{\rm s}} ~\frac{\delta\br{\chi \vek{\theta},\chi}}{a(\chi)}\;,
}

\noindent where $\delta$ is the three-dimensional matter density contrast, $\chi_{\rm s}$ is the comoving distance of the background galaxy which is acting as a source, and $a(\chi)$ is the cosmic scale factor at the comoving distance $\chi$ of $\delta$ which is acting as a lens. 

Equation (\ref{eq:d2k}) clearly shows that the contribution of the matter inhomogeneity $\delta$ at comoving distance $\chi_{i}$ to the cosmic shear signal of background galaxies can be considered as a function of the source distance $\chi_{\rm s}$, and this function is proportional to $1-{\chi_{i}}/{\chi_{\rm s}}$. The nulling technique takes advantage of this characteristic dependence on source distance $\chi_{\rm s}$ by constructing a weight function $T(\chi_i, \chi_{\rm s})$ such that the product of $T(\chi_i, \chi_{\rm s})$ and $1-{\chi_{i}}/{\chi_{\rm s}}$ has an average of zero on the range between $\chi_{i}$ and the comoving distance to the horizon $\chi_{\textrm{hor}}$:

\eq{
\label{eq:nc}
\int_{\chi_{i}}^{\chi_{\textrm{hor}}} d\chi_{\rm s} ~T(\chi_i, \chi_{\rm s})\, \br{1 - \frac{\chi_{i}}{\chi_{\rm s}}} = 0\;.
}

\noindent One then uses this weight function as a weight for integrating over the source distance:

\eq{
\label{eq:projectedkappanew}
\hat{\kappa}_{\rm G}(\chi_i, \vek{\theta}) := \int_{\chi_{i}}^{\chi_{\textrm{hor}}} d\chi_{\rm s}\; T(\chi_i, \chi_{\rm s})\; \kappa_{\rm G}(\vek{\theta},\chi_{\rm s})\;.
}

\noindent The resulting new measure of shear signal $\hat{\kappa}_{\rm G}(\chi_i, \vek{\theta}) $ is then free of contributions from the matter inhomogeneity at distance $\chi_{i}$. Note that although the weight function $T$ has two arguments $\chi_i$ and $\chi_{\rm s}$ here, we consider it as a function of $\chi_{\rm s}$ for a particular $\chi_{i}$. 

Consider a correlator $\ba{{\kappa}_{\rm obs}^{i}{\kappa}_{\rm obs}^{j}}$ with comoving distances $\chi_{i}< \chi_{j}$. With a similar decomposition as (\ref{eq:split}), it is straightforward to see that the GI term in it is $\ba{{\kappa}_{\rm I}^{i}{\kappa}_{\rm G}^{j}}$. The term $\ba{{\kappa}_{\rm G}^{i}{\kappa}_{\rm I}^{j}}$ vanishes since the lensing signal at $\chi_{i}$ is correlated only with matter with $\chi \le \chi_{i}$, whereas ${\kappa}_{\rm I}^{j}$ originates solely from physical processes happening at $\chi_{j}$. If we integrate $\ba{{\kappa}_{\rm I}^{i}{\kappa}_{\rm G}^{j}}$ over $\chi_j$ with a weight function that eliminates the contributions to ${\kappa}_{\rm G}^{j}$ by the matter inhomogeneity at distance $\chi_{i}$, this correlator will also vanish,
\eq{
\int_{\chi_{i}}^{\chi_{\textrm{hor}}} d\chi_j \; T(\chi_i, \chi_j)\; {\ba{{\kappa}_{\rm I}^{i}{\kappa}_{\rm G}^{j}}} = 0,}
\noindent since it is just the matter inhomogeneity at distance $\chi_{i}$ that gives rise to the correlation between ${\kappa}_{\rm I}^{i}$ and ${\kappa}_{\rm G}^{j}$. Thus, when we integrate over $\ba{{\kappa}_{\rm obs}^{i}{\kappa}_{\rm obs}^{j}}$ with the same weight function, the GI contamination in it will be ``nulled out''. Equation (\ref{eq:nc}) is the condition that the weight function $T$ should satisfy in order to ``null'' the intrinsic-shear alignment terms, so we call it ``the nulling condition''.

The same applies to three-point statistics. Consider a correlator $\ba{{\kappa}_{\rm obs}^{i}{\kappa}_{\rm obs}^{j}{\kappa}_{\rm obs}^{k}}$ with $\chi_{i}$ being the smallest comoving distance of the three. Both GII and GGI systematics contained in it also originate from the matter inhomogeneity at distance ${\chi}_{i}$. Typically, the generation of GII systematics requires that $\chi_{i} \approx \chi_{j} < \chi_{k}$, while the generation of GGI requires $\chi_{i} < \chi_{j}$ and $\chi_{i} < \chi_{k}$. For both cases, the dependence of GII or GGI systematics on $\chi_{k}$ is also just $1-{\chi_{i}}/{\chi_{k}}$. So new measures built as $\int_{\chi_{i}}^{\chi_{\textrm{hor}}} d\chi_{k} \; T(\chi_i, \chi_j, \chi_{k})\ba{{\kappa}_{\rm obs}^{i}{\kappa}_{\rm obs}^{j}{\kappa}_{\rm obs}^{k}}$ with $T$ satisfying the nulling condition for three-point statistics
\eq{
\int_{\chi_{i}}^{\chi_{\textrm{hor}}} d\chi_k ~T(\chi_i, \chi_j, \chi_k)\, \br{1 - \frac{\chi_{i}}{\chi_k}} = 0
}
\noindent will be free of both GII and GGI contamination. Again, $T(\chi_i, \chi_j, \chi_{k})$ here should be seen as a function of $\chi_{k}$ whose form depends on $\chi_i$ and $\chi_j$. 

Note that this method only depends on the characteristic redshift dependencies of the lensing signal and intrinsic-alignment signals, and is not limited to E-mode fields. This is a reassuring feature since while the $\kappa_{\rm G}$ field is a pure E-mode field to first order, the $\kappa_{\rm I}$ field can have a B-mode component. However, if parity-invariance is assumed, any correlation function which contains an odd number of B-mode shear components vanishes \citep{s03}, thus there should be no B-mode component in the GGI signal.

\subsection{Nulling formalism for lensing bispectrum tomography}
\label{sec:bispectomo}

Since the nulling technique relies on the distinct redshift dependence of the intrinsic-alignment signal, redshift information is crucial for it. With the help of infrared bands, forthcoming multicolor imaging surveys can provide rather accurate  photometric redshift information for the galaxies \citep[e.g.][]{abdalla08, bordoloi09}, allowing tomographic studies of cosmic shear statistics. We base our study on cosmic shear bispectrum tomography, and outline the corresponding formalism of the nulling technique in the following.  

Given the galaxy redshift probability distribution of redshift bin $i$ which we denote as $p^{(i)}_{\textrm{s}}(z)=p^{(i)}_{\textrm{s}}(\chi_{\rm s})\,d\chi_{\rm s}/dz\,$, one can define the average convergence field in redshift bin $i$ by integrating $\kappa(\vek{\theta},\chi_{\rm s})$ in (\ref{eq:d2k}) over $p^{(i)}_{\textrm{s}}(\chi_{\rm s})$. We turn to angular frequency space now and define
\eq{
\label{eq:kappai}
\tilde{\kappa}^{(i)}_{\rm G}(\vek{\ell}):=\int_0^{\chi_{\rm hor}}~d\chi_{\rm s}\, p_{\rm s}^{(i)}(\chi_{\textrm{s}}) ~\tilde{\kappa}_{\rm G}(\vek{\ell},\chi_{\rm s}) \;,
}
\noindent where $\tilde{\kappa}_{\rm G}(\vek{\ell},\chi_{\rm s})$ is the Fourier transform of $\kappa_{\rm G}(\vek{\theta},\chi_{\rm s})$. To better show the relation between $\tilde{\kappa}^{(i)}_{\rm G}(\vek{\ell})$ and three-dimensional matter inhomogeneity in Fourier space $\tilde{\delta}\br{k,\chi}$, one can write, basing on (\ref{eq:d2k}) and (\ref{eq:kappai}), 
 
\eq{ 
\label{eq:d2kappa}
\tilde{\kappa}^{(i)}_{\rm G}(\vek{\ell})=\int_0^{\chi_{\rm hor}}\!\!d\chi~ W^{(i)}(\chi)~ \tilde{\delta}\br{\vek{\ell}/ \chi,\chi}\;,
}
\noindent by defining a lensing weight function $W^{(i)}(\chi)$ as
\eq{
\label{eq:weight}
W^{(i)}(\chi) := \frac{3\, \Omega_{\textrm{m}}H_0^2 \,\chi}{2\, a(\chi)\, c^2} \int_{\chi}^{\chi_{\rm hor}}\!\!d\chi_{\textrm{s}}~ p_{\rm s}^{(i)}(\chi_{\textrm{s}}) \frac{\chi_{\textrm{s}}-\chi}{\chi_{\textrm{s}}}\;.
}

The tomographic lensing bispectrum is defined via
\eq{
\label{eq:fbisp}
\ba{\tilde{\kappa}_{\rm G}^{(i)}(\vek{\ell}_1)\tilde{\kappa}_{\rm G}^{(j)}(\vek{\ell}_2)
\tilde{\kappa}_{\rm G}^{(k)}(\vek{\ell}_3)}=(2\pi)^2
B^{(ijk)}_{\rm GGG}({\ell}_1,{\ell}_2,{\ell}_3)\;\delta_D(\vek{\ell}_1+\vek{\ell}_2+\vek{\ell}_3)\;,
}

\noindent where the Dirac delta function ensures that the bispectrum is defined only when $\vek{\ell}_1$, $\vek{\ell}_2$, and $\vek{\ell}_3$ form a triangle. This fact arises from statistical homogeneity, while that the bispectrum can be defined as a function independent of the directions of the angular frequency vectors arises from statistical isotropy. 

In a survey, the convergence field $\tilde{\kappa}_{\rm obs}$ is determined from the observed galaxy ellipticities, and the corresponding bispectrum $B_{\textrm{obs}}$ suffers from intrinsic-shear alignments. As we did with the three-point correlator in Sect.$\,$\ref{sec:introduction}, we separate the observed lensing bispectrum into the four terms:
\eq{
B_{\textrm{obs}}=B_{\textrm{GGG}}+B_{\textrm{GGI}}+B_{\textrm{GII}}+B_{\textrm{III}}\;.
}

\noindent Among them, $B_{\textrm {GGI}}$, $B_{\textrm{GII}}$ and $B_{\textrm{III}}$ can be linked to the convergence in a similar way as (\ref{eq:fbisp}), for example

\eq{
\label{eq:bGGI}
\ba{\tilde{\kappa}_{\rm I}^{(i)}(\vek{\ell}_1)\tilde{\kappa}_{\rm G}^{(j)}(\vek{\ell}_2)
\tilde{\kappa}_{\rm G}^{(k)}(\vek{\ell}_3)}=(2\pi)^2
B^{(ijk)}_{\rm GGI}({\ell}_1,{\ell}_2,{\ell}_3)\;\delta_D(\vek{\ell}_1+\vek{\ell}_2+\vek{\ell}_3)\;.
}
\noindent Here we assume disjunct redshift bins and let $i$ to be the redshift bin with the lowest redshift, so $\ba{\tilde{\kappa}_{\rm G}^{(i)}(\vek{\ell}_1)\tilde{\kappa}_{\rm I}^{(j)}(\vek{\ell}_2)\tilde{\kappa}_{\rm G}^{(k)}(\vek{\ell}_3)}$ and $\ba{\tilde{\kappa}_{\rm G}^{(i)}(\vek{\ell}_1)\tilde{\kappa}_{\rm G}^{(j)}(\vek{\ell}_2)\tilde{\kappa}_{\rm I}^{(k)}(\vek{\ell}_3)}$ both vanish due to the same reason as explained in Sect.$\,$\ref{sec:nulling} for the two-point statistics.

The purpose of the nulling technique is to filter $B_{\textrm{obs}}$ in such a way that the GGI term is strongly suppressed in comparison with the GGG term. The GII and III terms can be removed by ignoring the signal coming from bispectrum $B^{(ijk)}_{\rm obs}({\ell}_1,{\ell}_2,{\ell}_3)$ with two or three equal redshift bins.

To fulfill this purpose, we construct our new measures as
\eq{ 
\label{eq:Yij}
Y^{(ij)}({\ell}_1,{\ell}_2,{\ell}_3):=\sum_{k=i+1}^{N_z} T^{(ij)}(\chi_k)~B^{(ijk)}_{\textrm{obs}}({\ell}_1,{\ell}_2,{\ell}_3)\;\chi'_k\;\Delta z_k \;,
}

\noindent where $N_z$ is the total number of redshift bins, $\chi'_k$ is the derivative of comoving distance with respect to redshift, and $\Delta z_k$ is the width of redshift bin $k$. The weight function is written now as $T^{(ij)}(\chi_k)$ since $i$ and $j$ indicate two redshift bins, i.e. two populations of galaxies, rather than two comoving distances as in the previous subsection. The weight $T^{(ij)}$ is required to satisfy the nulling condition (\ref{eq:nc}) in its discretized form,
\eq{ 
\label{eq:nc2}
O^{(ij)}:=\sum_{k=i+1}^{N_z} T^{(ij)}(\chi_k) ~ \br{1-\frac{\chi_i}{\chi_k}} ~ \chi'_k \; \Delta z_k = 0\;,  
}

\noindent for all $j>i$. Here, $\chi_i$ and $\chi_k$ should be chosen such that they represent well the distance to redshift bins $i$ and $k$. In this paper we choose them to be the distances corresponding to the median redshift of the bin. The summation over index $k$ runs from $i+1$ rather than $i$ since we consider only bispectrum measures with $j > i$ and $k > i$ to avoid III and GII systematics. In this case $B^{(ijk)}_{\textrm{obs}}$ in (\ref{eq:Yij}) can be written as a sum of $B^{(ijk)}_{\textrm{GGG}}$ and $B^{(ijk)}_{\textrm{GGI}}$, and $Y^{(ij)}$ can be expressed as
\eq{
\label{eq:shownulling}
\begin{split}
& Y^{(ij)}({\ell}_1,{\ell}_2,{\ell}_3)=\sum_{k=i+1}^{N_z} T^{(ij)}(\chi_k)~B^{(ijk)}_{\textrm{GGG}}({\ell}_1,{\ell}_2,{\ell}_3)\;\chi'_k\;\Delta z_k \\
& + \sum_{k=i+1}^{N_z} T^{(ij)}(\chi_k)~B^{(ijk)}_{\textrm{GGI}}({\ell}_1,{\ell}_2,{\ell}_3)\;\chi'_k\;\Delta z_k\;.
\end{split}
}

Suppose one has infinitely many redshift bins, then the lensing signal in bin $k$ caused by the matter inhomogeneity in bin $i$ is exactly proportional to $1-\chi_i/\chi_k$, which means $B^{(ijk)}_{\textrm{GGI}}({\ell}_1,{\ell}_2,{\ell}_3)$ can be written as a product of $1-\chi_i/\chi_k$ and some function of the parameters other than $\chi_k$:
\eq{
\label{eq:GGIsplit}
B^{(ijk)}_{\textrm{GGI}}({\ell}_1,{\ell}_2,{\ell}_3) = {\cal F}(\chi_i, \chi_j, {\ell}_1,{\ell}_2,{\ell}_3)\, \br{1-\frac{\chi_i}{\chi_k}}\;.
}
\noindent Then we have
\eq{
\label{eq:tomonull}
\begin{split}
& \sum_{k=i+1}^{N_z} T^{(ij)}(\chi_k)~B^{(ijk)}_{\textrm{GGI}}({\ell}_1,{\ell}_2,{\ell}_3)\;\chi'_k\;\Delta z_k = \\
& {\cal F}(\chi_i, \chi_j, {\ell}_1,{\ell}_2,{\ell}_3) \, \sum_{k=i+1}^{N_z} T^{(ij)}(\chi_k)~\br{1-\frac{\chi_i}{\chi_k}} \;\chi'_k\;\Delta z_k = 0\;.
\end{split}
}
\noindent This suggests that only the GGG contribution is left in the nulled measure $Y^{(ij)}$, the GGI contribution has been ``nulled out'' due to the nulling condition. If only a limited number of redshift bins is available, (\ref{eq:GGIsplit}) holds only approximately, leading to a residual in (\ref{eq:tomonull}).

Since the nulling condition is the only condition that the weight $T^{(ij)}$ must satisfy in order to ``null'', there is much freedom in choosing the form of it. We would like to further specify its form such that it preserves as much Fisher information in $Y^{(ij)}$ as possible. The method we have adopted for the nulling weight construction will be detailed in Sect.$\,$\ref{sec:weight}.

Note that for each $(i,j)$ combination, one can in principle apply more than one nulling weight to the original bispectrum, and obtain more nulled measures. If one retains the condition of maximizing the Fisher information and demands that all the weight functions built for one $(i,j)$ combination are orthogonal to each other, one arrives at higher-order modes that have the second-most, third-most, etc., information content (higher-order weights, see JS08). The total number of such linearly independent nulled measures for a certain $(i,j)$ equals the possible values of $k \ge i+1$. In this schematic study we will only use the optimum, i.e. the first-order nulling weights. We will assess the information loss due to this limitation in Sect.$\,$\ref{sec:compression}.

\section{Modeling}
\label{sec:model}

\subsection{Survey characteristics}
\label{sec:setup}
We set up a fictitious survey with a survey size of $A$ = 4000 deg$^2$ which is similar to the survey size of DES. This can be easily scaled to any survey size using the proportionality of statistical errors to $A^{-1/2}$. We assume a galaxy intrinsic ellipticity dispersion $\sigma_{\epsilon}=\sigma(\epsilon_{\rm I}^{\rm ran})=0.35$. As galaxy redshift probability distribution we adopt the frequently used parameterization \citep{smail94},
\eq{
p_{\textrm{s}}(z) \propto \br{\frac{z}{z_0}}^{\alpha} \exp \bc{ -\br{\frac{z}{z_0}}^\beta},
}

\noindent and use $z_0 = 0.64$, $\alpha = 2$, $\beta = 1.5$. The distribution is cut at $z_{\textrm{max}} = 3$ and normalized to 1. The corresponding median redshift of this fictitious survey is $z_{\rm m}=0.9$, which is compatible to a survey like EUCLID. We adopt an average galaxy number density $\bar{n}_{\textrm{g}}=40~ {\textrm{arcmin}}^{-2}$ which is again EUCLID-like. 

Disjunct redshift bins without photo-z error are assumed, which means that the galaxy redshift probability distribution in redshift bin $i$ takes the form $p^{(i)}_{\textrm{s}}(z) \propto p_{\textrm{s}}(z)$ if and only if the redshift that corresponds to comoving distance $\chi_{\rm s}$ is within the boundaries of redshift bin $i$. A number of 10 redshift bins are used by default. The boundaries of the redshift bins are set such that each bin contains the same number of galaxies. 

We adopt 20 angular frequency bins spaced logarithmically between $\ell_{\textrm{min}}=50$ and $\ell_{\textrm{max}}=3000$, and denote the characteristic angular frequency of a bin as $\bar{\ell}$. Within this range the noise properties of the cosmic shear field are still not too far in the non-Gaussian regime, allowing a more realistic theoretical estimation of the bispectrum and its covariance. Whether this number of angular frequency bins can reconstruct the angular frequency dependence  of the bispectrum is tested, and 20 bins are found to be sufficient for our requirements on precision. This is also expected since the bispectrum is rather featureless as a function of angular frequency.

\subsection{Bispectrum and its covariance}

We show the modeling of $B_{\textrm{GGG}}$ and its covariance in this section. We will only consider the tomographic bispectrum at redshift bins satisfying $z_i<z_j$ and $z_i<z_k$, which already ensures an elimination of $B_{\textrm{III}}$ and $B_{\textrm{GII}}$ systematics in our case.

Applying Limber's equation, it can be shown that the tomographic convergence bispectrum can be written as a projection of the three-dimensional matter bispectrum $B_\delta\!\left({k_1},{k_2},{k_3}; \chi\right)$ (see e.g. TJ04):

\eq{
\label{eq:B_GGG}
\begin{split}
& B^{(ijk)}_{\rm GGG}(\bar{\ell}_1,\bar{\ell}_2,\bar{\ell}_3)= \\
& \int^{\chi_{\rm hor}}_0\!\!d\chi ~ \frac{W^{(i)}(\chi)W^{(j)}(\chi)W^{(k)}(\chi)}{\chi^{4}}~ B_\delta\!\left(\frac{\bar{\ell}_1}{\chi},\frac{\bar{\ell}_2}{\chi},\frac{\bar{\ell}_3}{\chi};
\chi\right)\;. 
\end{split}
}

To compute $B_{\delta}$, we employ the fitting formula by \citet{sco01}, which is based on hyper-extended perturbation theory \citep{sco99}. A comparison of this formula with the halo model results can be found in \citet{TJ03b,TJ03c}.

Estimating the bispectrum covariance is often done within a flat-sky spherical harmonic formalism \citep[Hu00 hereafter]{hu00}, which suffers - at least formally - from drawbacks since its basis functions $Y_{lm}$ are only defined for discrete angular frequency values and full sky coverage. In \citet{joachimi09b} another approach exclusively based on the two-dimensional Fourier formalism was constructed, which we use here.

The bispectrum covariance is a six-point correlation function which can be expanded into its connected parts as outlined in e.g. \citet{ber02a}. As argued in TJ04, for angular scales $\ell \le 3000$, the term which is a triple of two-point functions still dominates. Keeping only this term, the bispectrum covariance reads

\eq{
\label{eq:bicov}
\begin{split}
& {\textrm{Cov}} \br{B^{(ijk)}_{\rm GGG}({\bar{\ell}}_1,{\bar{\ell}}_2,{\bar{\ell}}_3),\;B^{(lmn)}_{\rm GGG}({\bar{\ell}}_4,{\bar{\ell}}_5,{\bar{\ell}}_6) } = \\
& \; \frac{(2\,\pi)^3}{A\; {\bar{\ell}_1}{\bar{\ell}_2}{\bar{\ell}_3} \; {\Delta \bar{\ell}_1}{\Delta \bar{\ell}_2}{\Delta \bar{\ell}_3}} \; \Lambda^{-1}\br{\bar{\ell}_1,\bar{\ell}_2,\bar{\ell}_3} \;   \\
& \times \br{ \bar{P}^{(il)}(\bar{\ell}_1)\bar{P}^{(jm)}(\bar{\ell}_2)\bar{P}^{(kn)}(\bar{\ell}_3)\;\delta_{\bar{\ell}_1 \bar{\ell}_4}\;\delta_{\bar{\ell}_2 \bar{\ell}_5}\;\delta_{\bar{\ell}_3 \bar{\ell}_6} + 5\, \mbox{perms.}},
\end{split}
}

\noindent in which $\Delta \bar{\ell}_i$ is the bin width of the angular frequency bin with typical value $\bar{\ell}_i$, and $\bar{P}^{(ij)}(\bar{\ell})$ is the observed power spectrum which contains the intrinsic ellipticity noise \citep[e.g.][]{kaiser92,hu99,JS08a}: 
\eq{
\label{eq:psshapenoise}
\bar{P}^{(ij)}(\bar{\ell}) = P^{(ij)}(\bar{\ell}) + \delta_{ij}\frac{\sigma^2_\epsilon}{2 \bar{n}_i}\;,
}
\noindent where $\bar{n}_i$ is the galaxy number density in redshift bin $i$. The term $\Lambda \br{\ell_1,\ell_2,\ell_3}$ is defined as
\eqa{
\label{eq:deflambda}
\begin{split}
\Lambda \br{\ell_1,\ell_2,\ell_3} \equiv \left\{  \begin{array}{ll} & \bc {\frac{1}{4} \sqrt{2 \ell_1^2 \ell_2^2 + 2 \ell_1^2 \ell_3^2 + 2 \ell_2^2 \ell_3^2 - \ell_1^4 - \ell_2^4 - \ell_3^4} }^{-1}   \\
& \hspace{2cm} \,\, \mbox{if}~~~ |\ell_1-\ell_2| < \ell_3 < \ell_1 + \ell_2\,, \\ 
& 0  \hspace{2cm} \mbox{else}.\\ \end{array} \right.
\end{split}
}

\noindent When $|\ell_1-\ell_2| < \ell_3 < \ell_1 + \ell_2$ is satisfied, $\Lambda^{-1} \br{\ell_1,\ell_2,\ell_3}$ is the area of a triangle with side lengths $\ell_1,\ell_2$ and $\ell_3$. We use the \citet{EH} transfer function to evaluate the linear three-dimensional matter power spectrum, and the \citet{SMITH} fitting function for the nonlinear power spectrum.  

Note that there is no intrinsic ellipticity noise in the observed bispectrum, since the galaxy intrinsic ellipticity distribution is assumed to be skewless. 

Under the assumptions of a compact survey geometry and scales much smaller than the extent of the survey area, (\ref{eq:bicov}) provides a bispectrum covariance that naturally incorporates the scaling with survey size, is not restricted to integer angular frequencies, and allows for any appropriate binning. In terms of Fisher information, the result given by this approach and the Hu00 one agree to high accuracy \citep{joachimi09b}.

\subsection{Toy intrinsic-shear alignment model}
In this section we present a toy model for generating GGI systematics. Since the physical generation of intrinsic-shear alignments concerns nonlinear growth of structure and complex astrophysical processes which are not easy to quantify, a realistic model is not yet available. Current simulations involving baryonic matter also have some way to go before they can simulate the generation of the GGI systematics reliably.

Up to now there has not been any attempt to measure GGI and GII in galaxy surveys. \citet{sembo08} studied these systematics using ray-tracing simulations. They provided fits in real space to projected GII and GGI signals, but the results are still too crude to lead to sufficient constraints on an intrinsic-shear alignment model. 

This situation emphasizes the importance of a method intended to control intrinsic-shear alignment to be model-independent, especially at the three-point level. Since this is the case for the nulling technique, for this work we only require a simple model for $B_{\textrm {GGI}}^{(ijk)}$ which satisfies the characteristic redshift dependence and leads to a reasonable bias. 

Based on the observation that the lensing bispectrum expression (\ref{eq:B_GGG}) comes directly from (\ref{eq:d2kappa}) and the definition of the tomography bispectrum (\ref{eq:fbisp}), we link $B_{\textrm {GGI}}^{(ijk)}$ also to a three-dimensional bispectrum $B_{\delta_I\delta\delta}$ via

\eq{
\label{eq:ggiia}
\begin{split}
& B_{\textrm {GGI}}^{(ijk)}(\bar{\ell}_1,\bar{\ell}_2,\bar{\ell}_3)= \\
& \int^{\chi_{\rm hor}}_0\!\!d\chi ~\frac{p_{\rm s}^{(i)}(\chi) W^{(j)}(\chi)W^{(k)}(\chi)}{\chi^{4}}~ B_{\delta_I\delta\delta}\br{\frac{\bar{\ell}_1}{\chi},\frac{\bar{\ell}_2}{\chi},\frac{\bar{\ell}_3}{\chi};\chi}.
\end{split} 
}

\noindent Similar to $B_{\delta}(k_1,k_2,k_3)$ which is given by
\eq{
\begin{split}
&\ba{\tilde{\delta}(\vek{k_1},\chi)\tilde{\delta}(\vek{k_2},\chi)\tilde{\delta}(\vek{k_3},\chi)}=\\
&\hspace{2cm} (2\pi)^3\,\delta_D(\vek{k_1}+\vek{k_2}+\vek{k_3})\,B_{\delta}(k_1,k_2,k_3;\chi),
\end{split} 
}
\noindent $B_{\delta_I\delta\delta}$ is defined via
\eq{
\begin{split}
& \ba{\tilde{\delta}_I(\vek{k_1},\chi)\tilde{\delta}(\vek{k_2},\chi)\tilde{\delta}(\vek{k_3},\chi)}=\\
&\hspace{2cm} (2\pi)^3\,\delta_D(\vek{k_1}+\vek{k_2}+\vek{k_3})\,B_{\delta_I\delta\delta}(k_1,k_2,k_3;\chi),
\end{split} 
}
\noindent where $\tilde{\delta}_{\rm I}(\vek{k})$ is the three-dimensional density field which is responsible for the intrinsic alignment, and it satisfies
\eq{ 
\label{eq:d2kappa_I}
\tilde{\kappa}^{(i)}_{\rm I}(\vek{\ell})=\int_0^{\chi_{\rm hor}}\!\!d\chi~ p_{\rm s}^{(i)}(\chi)~ \tilde{\delta}_{\rm I}\br{\frac{\vek{\ell}}{\chi},\chi}\;,
}
\noindent The definition of both $\tilde{\kappa}^{(i)}_{\rm I}$ and $\tilde{\delta}_{\rm I}$ originates from the deterministic part of galaxy intrinsic ellipticity $\epsilon_{\rm I}^{\rm det}$. We have assume the existence of these underlying smooth fields. Similar quantities have been defined in \citet{joachimi09c}, see also \citet{hirata04} and \citet{schneider09}. We would like to point out again that, although we introduce these quantities for the clarity of our model, we do not need them for the main purpose of this paper. What we need to model is the projected GGI bispectrum $B_{\textrm {GGI}}^{(ijk)}$.

Note that in (\ref{eq:ggiia}), the weight for the lowest redshift bin $i$ is the source distribution function $p_{\rm s}^{(i)}$ which is zero outside redshift bin $i$, rather than the lensing weight $W^{(i)}$ which is a much broader function. Since $\tilde{\kappa}_{\rm I}^{(i)}$ depends only on physical processes at redshift bin $i$ and is inferred from ellipticity measurements in this bin, and $\tilde{\kappa}_{\rm G}^{(j)}$ is linked to the three-dimensional matter density through the lensing weight $W^{(j)}$, this assignment of weight functions will ensure the correct redshift dependence of $B_{\textrm {GGI}}^{(ijk)}$. 

When the redshift bins are not disjunct, however, the intrinsic alignment signal can no longer be associated with bin $i$. There will be two permutations in both the left-hand side of (\ref{eq:bGGI}) and the right-hand side of (\ref{eq:ggiia}), similar to the two-point case, e.g. Eq.\,11 in \citet{hirata04}.  

\begin{figure*}[t]
\centering
\includegraphics[width=18cm]{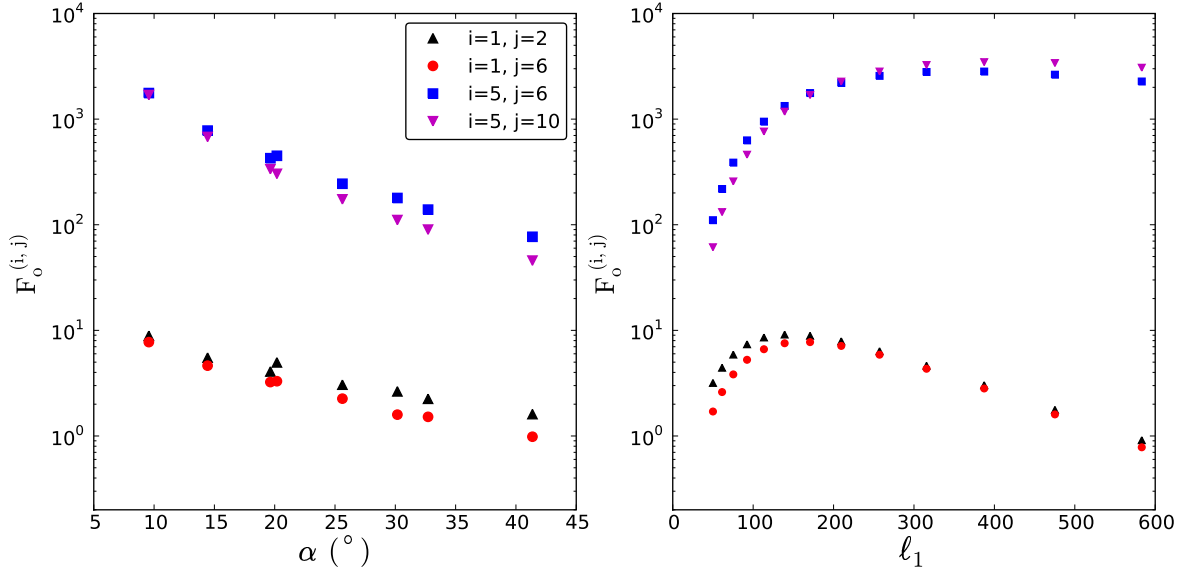}
\caption{Distribution of the nulled Fisher information as defined in (\ref{eq:foij}) per $(\bar{\ell}_1, \bar{\ell}_2, \bar{\ell}_3)$ bin and per redshift bin combination among different angular frequency triangle shapes and sizes. Results for four redshift bin combinations $(i,j)$ are presented. 
\textit{Left panel}: Distribution of the nulled Fisher information among different triangle configurations. We consider triangles with the common shortest side length $\bar{\ell}_1=171$ which corresponds to the 7th angular frequency bin. Due to our logarithmic binning and the constraint that the three side lengths must be able to form a triangle, only 8 such triangle configurations exist. Plotted is the nulled Fisher information contained in these 8 triangles against $\alpha$, which is the angle opposite to the shortest side length in that triangle. Smaller $\alpha$ correspond to more elongated triangles, and larger $\alpha$ correspond to almost equilateral triangles. 
\textit{Right panel}: Distribution of the nulled Fisher information contributed by each $(\bar{\ell}_1, \bar{\ell}_2, \bar{\ell}_3)$ bin over different triangle sizes. A fixed triangle shape with $\bar{\ell}_1:\bar{\ell}_2:\bar{\ell}_3=1:3.64:4.52$ (corresponds to the leftmost points in the left panel) is chosen. The nulled Fisher information contained in one $(\bar{\ell}_1, \bar{\ell}_2, \bar{\ell}_3)$ bin is plotted against the shortest side length $\bar{\ell}_1$ of each triangle.} 
\label{fig: loptimal}
\end{figure*}

The modeling of $B_{\delta_I\delta\delta}$ is then a pure matter of choice. We build a simple three-dimensional GGI bispectrum with power-law dependence on both redshift $z$ and spatial frequency $k$:
\begin{equation}
\label{eq:IAmodel}
\begin{split}
& B_{\delta_I\delta\delta}\br{{k_1},{k_2},{k_3};\chi} := - {\cal A} \, B_{\delta}\,(k_{\textrm {ref}},k_{\textrm {ref}},k_{\textrm {ref}};\chi(z_{\textrm {med}}))\, \br{\frac{1+z}{1+z_{\textrm{med}}}}^{r-2}\\ 
& \times \;\bc{{\br{\frac{k_1}{k_{\textrm {ref}}}}}^{2(s-2)} + {\br{\frac{k_2}{k_{\textrm {ref}}}}}^{2(s-2)} + {\br{\frac{k_3}{k_{\textrm {ref}}}}}^{2(s-2)}}\;,
\end{split}
\end{equation}

\noindent where $z_{\textrm {med}}$ is the median redshift of the whole survey, and ${\cal A}$, $k_{\textrm {ref}}$, $r$, $s$ are free parameters. Among them the parameter $k_{\textrm {ref}}$ is designed to be a characteristic wave number, whose value we set to be a weakly nonlinear scale of $10\;h{\rm Mpc}^{-1}$ here. The minus sign ensures that the contamination of GGI systematics leads to an underestimation of the GGG signal, as found by \citet{sembo08}. 

Little is known about the redshift and angular scale dependence of $B_{\delta_I\delta\delta}$. However one can roughly estimate how it compares to the $B_{\delta\delta\delta}$ signal. 
A linear alignment model suggests $\delta_I \propto \delta_{\rm lin}\,\bar{\rho}(z)/((1+z)\,D_{+}(z))$ \citep[see e.g.][]{hirata04}, in which $\bar{\rho}(z)$ is the mean density of the universe, ${D_{+}}(z)$ is the growth factor, and $\delta_{\rm{lin}}$ is the linear matter density contrast. Thus we have, very roughly, $\delta_I \propto (1+z)^3\,\delta_{\rm lin}$ which suggests $B_{\delta_I\delta\delta} \propto (1+z)^3\,B_{\delta\delta\delta}$.
The linear alignment model assumes that the intrinsic alignment is linearly related to the local tidal gravitational field \citep[e.g.][]{catelan01,hirata04}. If this holds true, we also expect $B_{\delta_I\delta\delta}$ to have a stronger angular scale dependence than $B_{\delta\delta\delta}$ since tidal gravitational interaction follows the inverse cube law rather than the inverse square law which gravity itself follows.
For a $\Lambda$CDM model, in the weakly nonlinear regime where perturbation theory holds, the dependence of $B_{\delta\delta\delta}$ on $(1+z)$ has a negative power shallower than $-4$, and the dependence on $k$ has a power of around $-2$. In this paper we choose $r=0$, $s=1$ as default. We also study the cases of $r=-2$, $r=2$, and $s=0$ whose results will be shown in Fig.$\,$\ref{fig:nz} below. 

\begin{figure*}[t]
\centering
\includegraphics[width=18cm]{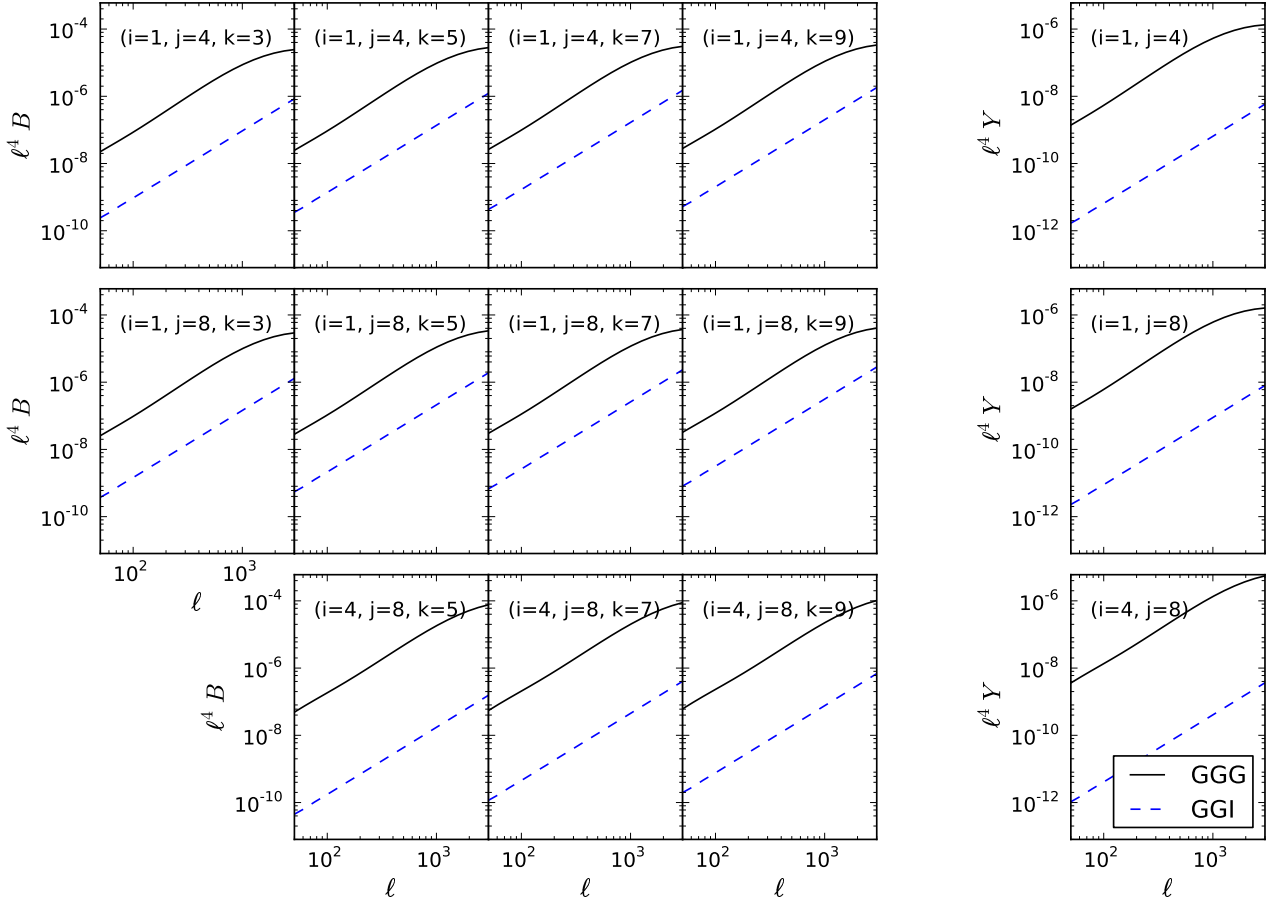}
\caption{Tomographic convergence bispectrum (GGG, solid curves) and intrinsic-shear alignment (GGI, dashed curves) for equilateral triangles are plotted against triangle side length. Measures both before (left panel) and after (right panel) applying the nulling technique ($B$ and $Y$ respectively) are shown for three typical redshift bin $(i,j)$ combinations in the three rows.}
\label{fig:ggiggg}
\end{figure*}

As for the amplitude ${\cal A}$ of the GGI signal, the only direct study up to now is \citet{sembo08}, which suggests an overall GGI/GGG ratio of 10$\,\%$ for a $z_{\rm m} = 0.7$ survey for elliptical galaxies and few percent for a mixed sample of elliptical and spiral galaxies. In this paper we adjust ${\cal A}$ such that the amplitude of the tomographic GGI bispectrum is limited to be within 10$\,\%$ of the amplitude of the lensing GGG signal, i.e. GGI/GGG$\lesssim 10\,\%$ at redshift bin combinations with $z_i \ll z_j$ and $z_i \ll z_k$ where the GGI signal is expected to be most significant. This leads to a relatively modest overall GGI/GGG ratio at percent level. We will show examples of the generated GGI and GGG signals in Fig.$\,$\ref{fig:ggiggg}.
As an order-of-magnitude estimate, one can also relate the GGI/GGG ratio to that of GI/GG by expanding three-point signals to couples of two-point signals using perturbation theory, in analogy to the \citet{sco01} fitting formula. For the case of $z_i\ll z_j \approx z_k$, the leading order terms would give that the GGI/GGG ratio approximates that of GI/GG evaluated at redshifts $z_i$ and $z_j$. This suggests that our adopted GGI/GGG ratio is also consistent with available observational studies of the GI signal \citep{mandel06, mandel09, hirata07, fu08, okumura09a, okumura09b}, although the results of these studies vary a lot according to different median redshift, color and luminosity of the selected galaxy sample.

\section{Construction of nulling weights }
\label{sec:weight}

As mentioned in Sect.$\,$\ref{sec:bispectomo}, we would like to construct a single first-order weight function $T^{(ij)}(\chi)$ for each $(i,j)$ combination which preserves the maximum of information. This can be seen as a constrained optimization problem. The constraining condition here is the nulling condition and the quantity to be optimized is the Fisher information after nulling. In JS08, several practical methods were developed to solve this optimization problem at the two-point level, and very good agreement was found among the different methods. 

We adopt the simplified analytical approach as described in JS08, and reformulate it for three-point statistics here. For convenience we introduce the following notations: 

\noindent the bispectrum covariance matrix ${\rm \textbf CovB}$, whose elements are
\eq{
{\rm CovB}({^{ijk}_{\bar{\ell}_1,\bar{\ell}_2,\bar{\ell}_3}};{^{lmn}_{\bar{\ell}_4,\bar{\ell}_5,\bar{\ell}_6}}) := {\textrm{Cov}} \br{B_{\rm GGG}^{(ijk)}(\bar{\ell}_1,\bar{\ell}_2,\bar{\ell}_3),\;B_{\rm GGG}^{(lmn)}(\bar{\ell}_4,\bar{\ell}_5,\bar{\ell}_6) }\;;
}

\noindent the covariance matrix ${\rm \textbf CovY}$ of the nulled bispectra $Y$, whose elements are

\eq{
\label{eq:cov_Y}
\begin{split}
& {\rm CovY}({^{ij}_{\bar{\ell}_1,\bar{\ell}_2,\bar{\ell}_3}};{^{lm}_{\bar{\ell}_4,\bar{\ell}_5,\bar{\ell}_6}}) := \textrm{Cov}\br{Y^{(ij)}(\bar{\ell}_1,\bar{\ell}_2,\bar{\ell}_3),Y^{(lm)}(\bar{\ell}_4,\bar{\ell}_5,\bar{\ell}_6)} \\
& = \sum_{k= \atop i+1}^{N_z} \sum_{n= \atop l+1}^{N_z} \textrm{Cov} \br{B_{\rm GGG}^{(ijk)}(\bar{\ell}_1,\bar{\ell}_2,\bar{\ell}_3),B_{\rm GGG}^{(lmn)}(\bar{\ell}_4,\bar{\ell}_5,\bar{\ell}_6)} \\
& \times \, T^{(ij)}\br{\chi_{k}} T^{(lm)} \br{\chi_{n}} \chi'_{k} \;\chi'_{n} \;\Delta z_{k} \;\Delta z_{n}\; ;
\end{split}
}

\noindent a vector ${\rm \textbf B_{\textbf ,\,{\mu} }}$ whose elements are partial derivatives of the bispectrum with respect to the cosmological parameter $p_{\mu}$ 
\eq{
{ B_{ ,\,\mu}}({^{ijk}_{\bar{\ell}_1,\bar{\ell}_2,\bar{\ell}_3}}) := \frac{\partial B_{\rm GGG}^{(ijk)}(\bar{\ell}_1,\bar{\ell}_2,\bar{\ell}_3)}{\partial p_{\mu}}\;;
}

\noindent and a corresponding vector ${\rm \textbf Y_{\textbf ,\,\mu}}$ for nulled bispectra $Y$, whose elements are 
\eq{
{Y_{ ,\,\mu}}({^{ij}_{\bar{\ell}_1,\bar{\ell}_2,\bar{\ell}_3}}) := \frac{\partial Y^{(ij)}(\bar{\ell}_1,\bar{\ell}_2,\bar{\ell}_3)}{\partial p_{\mu}}\;.
}

Then the Fisher information matrix from the original bispectra can be written as (following TJ04) 

\eq{
\label{eq:F_i}
{\rm  F^{\rm i}_{\mu\nu}} = {\rm \textbf B_{\textbf ,\,\mu}}\, {\rm \textbf CovB}^{-1}\, {\rm \textbf B_{\textbf ,\,\nu}}\;\;,
}

\noindent and that from the nulled bispectra can be written as 

\eq{
\label{eq:F_f}
{\rm  F^{\rm f}_{\mu\nu}} = {\rm \textbf Y_{\textbf ,\,\mu}}\, {\rm \textbf CovY}^{-1}\, {\rm \textbf Y_{\textbf ,\,\nu}}\;\;.
} 

\noindent Here the matrix multiplication is a summation of all possible angular frequency combinations $(\bar{\ell}_1, \bar{\ell}_2, \bar{\ell}_3)$ and redshift bin combinations, ${(ijk)}$ for the original bispectra and ${(ij)}$ for the nulled bispectra. In (\ref{eq:F_i}) and (\ref{eq:F_f}), ${\rm \textbf CovB}^{-1}$ and ${\rm \textbf CovY}^{-1}$ indicate the inverse of the covariance matrix. When the covariance is approximated by triples of power spectra, the covariance between two different angular frequency combinations $(\bar{\ell}_1, \bar{\ell}_2, \bar{\ell}_3)\neq(\bar{\ell}_4,\bar{\ell}_5,\bar{\ell}_6)$ is zero, see (\ref{eq:bicov}), which means that the covariance matrix is block diagonal. In this case the matrix inversion can be done separately for each block specified by an angular frequency combination $(\bar{\ell}_1, \bar{\ell}_2, \bar{\ell}_3)$. 

According to the idea of the simplified analytical approach, we consider the Fisher information on one cosmological parameter contained in bispectrum measures $B_{\rm{GGG}}^{(ijk)}(\bar{\ell}_1, \bar{\ell}_2, \bar{\ell}_3)$ with a single $(\bar{\ell}_1, \bar{\ell}_2, \bar{\ell}_3)$ combination and with redshift bin $(i,j,k)$ combinations having common $(i,j)$ indices. For every $(i,j)$ combination we build nulling weights $T^{(ij)}$ which maximizes the nulled Fisher matrix using the method of Lagrange multipliers. Since here the nulled Fisher matrix receives contribution only from certain angular frequency and redshift combinations, we denote it as $F_{\rm o}^{(ij)}$ to avoid ambiguity. $F_{\rm o}^{(ij)}$ has only one component since only one cosmological parameter is taken into consideration. As only a single $(\bar{\ell}_1, \bar{\ell}_2, \bar{\ell}_3)$ combination is involved, we will omit the $\bar{\ell}$-dependence in all variables in the rest of this subsection to keep a compact form.

Again for notational simplicity, we follow JS08 and introduce a vector notation as follows. For each $(i,j)$ in consideration, let the values of the weights $T^{(ij)}(\chi_k)$ form a vector $\vek{T}=T_k$, and define another vector $\vek{\rho}$ and a matrix $\bar{\bf{C}}$ with elements

\eq{
\rho_{k}:= { B_{,\,\mu}}({^{ijk}}) \, \chi'_{k}\;\Delta z_{k} \;,
}

\eq{
\bar{C}_{kn} := {\rm CovB}({^{ijk}};{^{ijn}}) \;\chi'_{k} \;\chi'_{n} \;\Delta z_{k} \;\Delta z_{n}\;.
}

\noindent Thus $F_{\textrm{o}}^{(ij)}$ can be expressed, according to (\ref{eq:F_f}), as

\eq{
\label{eq:foij}
F_{\textrm{o}}^{(ij)} := {\rm \textbf Y_{\textbf ,\,\mu}}({^{ij}})\,\;{{\rm \textbf CovY}^{-1}({^{ij}};{^{ij}})}\,\;{\rm \textbf Y_{\textbf ,\,\mu}}({^{ij}}) = \frac{\br{\vek{T} \cdot \vek{\rho}}^2} {\vek{T}^\tau \bar{\bf{C}} \vek{T}}.
}

\noindent We further define a vector $\vek{f}$ with elements

\eq{
f_{k} = \br{1-\frac{\chi_{i}}{\chi_{k}}}\;\chi'_{k}\;\Delta z_{k} 
}

\noindent to write the nulling condition (\ref{eq:nc2}) as
\eq{
O^{(ij)} = \vek{T} \cdot \vek{f} = 0 \;.
}

The problem of finding nulling weights $\vek{T}$ which maximize $F^{(ij)}_o$ under the constraint given by the nulling condition can be solved with the method of Lagrange multipliers by defining a function

\eq{
\label{eq:maxi}
G := F_{\textrm{o}}^{(ij)} + \lambda O^{(ij)} =  \frac{\br{\vek{T} \cdot \vek{\rho}}^2} {\vek{T}^\tau \bar{\bf{C}} \vek{T}} + \lambda \vek{T}\cdot\vek{f}\;
}

\noindent with $\lambda$ being the Lagrange multiplier, and setting the gradient of $G$ with respect to $\vek{T}$ to zero,

\eq{
\nabla_T\; G = 2 \vek{\rho}\; \frac{\br{\vek{T} \cdot \vek{\rho}}}{\vek{T}^\tau \bar{\bf{C}} \vek{T}} - 2\; \bar{\bf{C}} \vek{T} \;\br{ \frac{\br{\vek{T} \cdot \vek{\rho}}}{\vek{T}^\tau \bar{\bf{C}} \vek{T}} }^2 + \lambda \vek{f} = 0\;.
}

\noindent The solution to this equation is (for more details see JS08)

\eq{
\vek{T} = {\cal N} \bc{\bar{\bf{C}}^{-1} \vek{\rho} - \frac{\vek{f}^\tau \bar{\bf{C}}^{-1} \vek{\rho}}{\vek{f}^\tau \bar{\bf{C}}^{-1} \vek{f}}\; \bar{\bf{C}}^{-1} \vek{f} }\;,
}

\noindent with the normalization $\cal N$ adjusted to give ${|T|}^2=1$.

Apparently the thus constructed nulling weights depend on which $(\bar{\ell}_1, \bar{\ell}_2, \bar{\ell}_3)$ combination is considered and with respect to which cosmological parameter we optimize the information content. In this paper the default cosmological parameter to optimize is $\Omega_{\rm m}$, and we choose for each $(i,j)$ combination the $(\bar{\ell}_1, \bar{\ell}_2, \bar{\ell}_3)$ combination which maximizes $F_{\rm o}^{(ij)}$. However one needs to be aware that this serves only as a clear choice of a $(\bar{\ell}_1, \bar{\ell}_2, \bar{\ell}_3)$ combination and is not necessarily the best in terms of information preservation considering all angular frequency bins and all cosmological parameters.

To show which triangle shapes and sizes contain more information, we plot $F_{\rm o}^{(ij)}$ against the $(\bar{\ell}_1, \bar{\ell}_2, \bar{\ell}_3)$ triangle shape and size for four typical $(i,j)$ combinations in Fig.$\,$\ref{fig: loptimal}. In the left panel, the nulled information $F_{\rm o}^{(ij)}$ contained in different triangles with a common shortest side length $\bar{\ell}_1=171$ is plotted against $\alpha$, which is the angle opposite to $\bar{\ell}_1$. Due to our logarithmic binning in angular frequency, only eight $(\bar{\ell}_1, \bar{\ell}_2, \bar{\ell}_3)$ combinations with $\bar{\ell}_1=171$ can form triangles. One sees that the more elongated triangles (small $\alpha$) contain much more Fisher information than the almost equilateral triangles (large $\alpha$). The small separation between the 3rd and the 4th points from the left is caused by the degeneracy of different triangle shapes with respect to $\alpha$, e.g. two equal and very long side lengths can result in the same value of $\alpha$ as two shorter side lengths with a length difference close to the length of the shortest side length. The right panel shows the distribution of the Fisher information contained in one $(\bar{\ell}_1, \bar{\ell}_2, \bar{\ell}_3)$ bin over the triangle size. When the redshift in consideration is higher, the peak of the information distribution moves to higher angular frequencies. The figure suggests that most information comes from high redshifts and small angular scales. 
  
\begin{figure}[t]
\centering
\includegraphics[width=9cm]{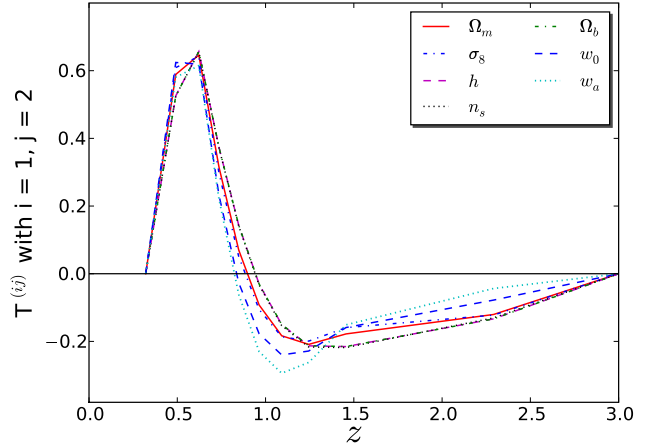}
\caption{Nulling weights $T^{(ij)}$ for redshift bins $i=1, j=2$ are plotted against the redshift value of the third redshift index $k$. Remarkable consistency is found between nulling weights optimized on different parameters, shown with different line styles.}
\label{fig: poptimal}
\end{figure}

To explore the sensitivity of nulling weights on the choice of the cosmological parameter, we construct seven sets of weight functions, each optimizing the information content in terms of one parameter. For all $(i,j)$ combinations we find that the nulling weights are not very sensitive to the choice of parameter. As an example, the weights for $(i,j) = (1,2)$ are shown in Fig.$\,$\ref{fig: poptimal}. This result is rather surprising at first sight, since for different parameters the distribution of information (contained in the bispectrum) over redshift bins is quite different. However, such insensitivity suggests that the shapes of nulling weights are already strongly constrained under our construction scheme. One constraint is, evidently, the nulling condition. Moreover, considering the fact that we optimize the nulling weights for each $(i,j)$ combination with respect to the information content they preserve, we have already required the shapes of these first order nulling weights to be as smooth as possible.

The fact that these two conditions have already imposed strong constraints on the nulling weights also suggests that nulling weights can be robustly and efficiently constructed, i.e. it is not critical to construct the ``best'' nulling weights.

\section{Performance of the nulling technique}
\subsection{GGI/GGG ratio}
\label{sec:ratio}
What the nulling technique ``nulls'' is the GGI signal $B_{\rm GGI}$, so the GGI/GGG ratio is the most direct quantification of its performance. We plot the modeled GGI and GGG bispectra before and after nulling in Fig.$\,$\ref{fig:ggiggg}. The original GGI signal is shown in the left panels by dashed lines. For comparison the GGG signals are shown as solid curves. The results are shown for equilateral triangle configurations for the convenience of presenting. One sees that when the redshift bin number $j$ and/or $k$ increase, the changes in GGG and GGI signals are different, which shows the expected different redshift dependence. For all redshift bin combinations the GGI signal is modeled to be subdominant to the GGG signal. In the nulled measures shown in the right panels, the GGI/GGG ratio is suppressed by a factor of 10 over all angular scales, which reflects the success of the nulling technique.

\subsection{Information loss and downweighting of systematics}
We further evaluate the performance of the nulling technique by looking at the constraining power of cosmic shear bispectrum tomography on cosmological parameters, as well as the biases caused by the GGI systematics before and after nulling.
 
The full characterization of the bispectrum involves three angular frequency vectors which form a triangle. In some works  concerning three-point statistics, only equilateral triangle configurations i.e. $\ell_1=\ell_2=\ell_3=\ell$ are used for simplicity reasons \citep[e.g.][]{pires09}. But as several authors have pointed out \citep[e.g.][]{kilbinger05,berge09}, only a low percentage of information is contained in equilateral triangles. Thus, to calculate the full information content, we use general triangle configurations but limit our calculation to triangles with three different side lengths, again for reasons of simplicity (for details see Appendix\,\ref{app:1} ).

We will use the figure of merit (FoM, \citealp{TaskForce06}) to quantify the goodness of parameter constraints. Here the FoM for constraints in the parameter plane $p_\alpha-p_\beta$ is defined to be proportional to the inverse of the area of the parameter constraint ellipses: 
\eq{
\label{eq:fom}
\rm{FoM}(p_\alpha,p_\beta) \equiv \br{(\rm \textbf F^{-1})_{\alpha \alpha}(\rm \textbf F^{-1})_{\beta \beta}-(\rm \textbf F^{-1})_{\alpha \beta}^{2}}^{-\frac{1}{2}}. 
}

To compute biases, we adopt a method based on a simple extension of the Fisher matrix formalism \citep[e.g.][]{hut06,amara08}. Then one needs to define a bias vector $\mathbb{B}^{\textrm{GGI}}$ which in our case reads: 

\eq{
\label{eq:Bsys_i}
{\mathbb B}_{\nu, \textrm{i}}^{\textrm{GGI}}={\rm \textbf B_{\rm GGI }}\, {\rm \textbf CovB}^{-1}\, {\rm \textbf B_{\textbf ,\,\nu}}\;\;,
}

\eq{
\label{eq:Bsys_o}
{\mathbb B}_{\nu, \textrm{f}}^{\textrm{GGI}}={\rm \textbf Y_{\rm GGI }}\, {\rm \textbf CovY}^{-1}\, {\rm \textbf Y_{\textbf ,\,\nu}}\;\;,
}

\noindent with

\eq{
{\rm \textbf  B_{\rm GGI }}({^{ijk}_{\bar{\ell}_1,\bar{\ell}_2,\bar{\ell}_3}}) :=  B_{\rm GGI}^{(ijk)}(\bar{\ell}_1,\bar{\ell}_2,\bar{\ell}_3)\;\;,
}

\eq{
{\rm \textbf  Y_{\rm GGI }}({^{ij}_{\bar{\ell}_1,\bar{\ell}_2,\bar{\ell}_3}}) :=  Y_{\rm GGI}^{(ij)}(\bar{\ell}_1,\bar{\ell}_2,\bar{\ell}_3)\;\;.
}

\noindent The bias of the parameter estimator $\hat{p}_\mu$ is given by the difference between its ensemble average and the fiducial value of the parameter $p_{\mu}^{\rm fid}$:
\eq{
\label{eq:bias}
b_{\mu}= \langle \hat{p}_\mu \rangle -  p_{\mu}^{\rm fid}  =\sum_{\nu} \br{\rm \textbf F^{-1}}_{\mu\nu} \mathbb{B}_{\nu}^{\textrm{GGI}} \;.
}

\begin{figure*}[t]
\begin{minipage}[c]{.75\textwidth}
\includegraphics[width=13cm]{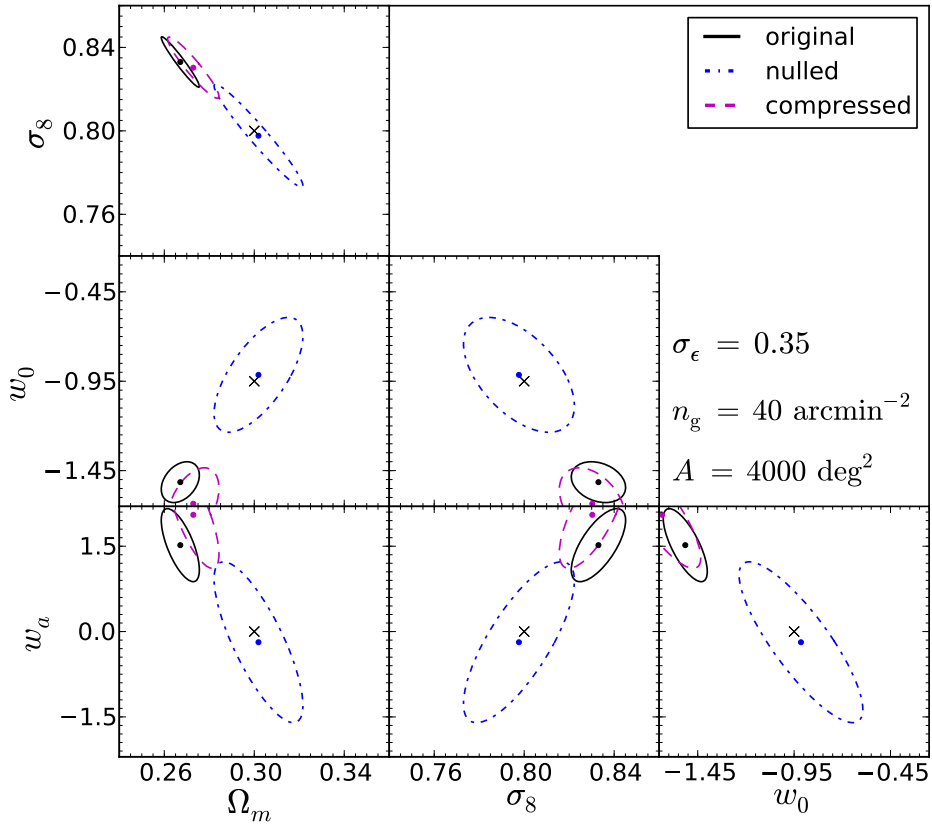}
\end{minipage}
\begin{minipage}[c]{.22\textwidth}
\caption{Projected 1-sigma (68$\,\%$ CL) parameter constraints from cosmic shear bispectrum tomography. Hidden parameters are marginalized over. The black solid and blue dash-dotted ellipses correspond to the original constraints and those after nulling, respectively. The black cross in the center of each panel represents the fiducial values adopted for the parameters, and the distance from the center of one ellipse to the black cross reflects the bias caused by intrinsic-alignment GGI systematics on the corresponding parameter. As nulling can be seen as a linear data compression under the constraint of the nulling condition, we also plot the constraints and biases after an unconditioned linear data compression as magenta dashed ellipses for comparison (see Sect.$\,$\ref{sec:compression}). } 
\label{fig:para}
\end{minipage}
\end{figure*}

The information content before and after nulling can be seen in Fig.$\,$\ref{fig:para}. On the cost of increasing the error on each parameter to about twice its original value, GGI systematics are reduced to be within the original statistical error. The relative information loss in terms of FoM can be found in Table$\,$\ref{tab:FoM}. The constraints shown in Fig.$\,$\ref{fig:para} do not represent the best constraints obtainable from a cosmic shear bispectrum analysis since we consider only the triangles with angular scale $\bar{\ell}_1 \ne \bar{\ell}_2 \ne \bar{\ell}_3$. Also note that the nulling technique can in principle remove the GGI systematics completely. But as shown in Fig.$\,$\ref{fig:para}, the systematics still cause some residual biases on cosmological parameters after nulling, due to the finite number of redshift bins. The GGI systematics will be reduced to a lower level when more redshift bins are available. We will discuss this further in the following subsection.

\subsection{How many redshift bins are needed?}

Analyzing the cosmic shear signal in a tomographic way was originally meant to maximize the information. For this purpose alone, a crude redshift binning will suffice \citep{hu99}. However, to control intrinsic-shear alignment, which is a redshift-dependent effect, much more detailed redshift information is required \citep[e.g.][]{king02,bridle07b,JS08b}. Thus, for a method intended to eliminate intrinsic-shear alignment, it is necessary to show its requirement on the redshift precision. In the case of nulling, detailed redshift information is not only needed for the method to be able to eliminate the bias, but also for the preservation of a reasonably large amount of information through the nulling process. JS08 examined the number of redshift bins required for the nulling technique in the two-point case, and showed that 10 redshift bins already ensure that parameters are still well-constrained after nulling.

\begin{figure}[h]
\centering
\includegraphics[width=9cm]{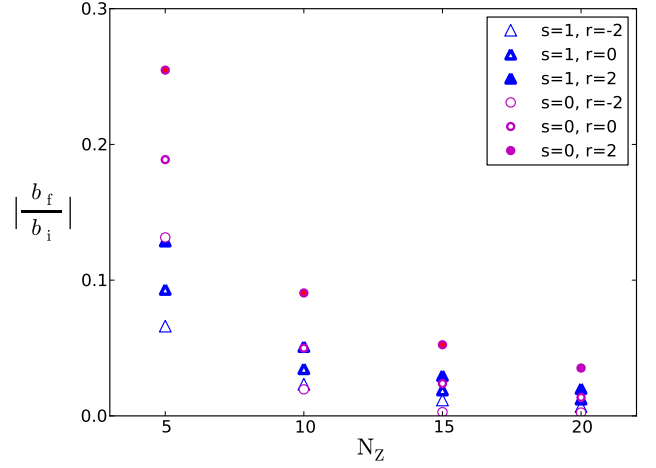}
\caption{ Ratio of the nulled and the original biases for cosmological parameter $\Omega_{\textrm m}$ as a function of number of redshift bins $N_z$. Results for different GGI models are shown. Parameter $s$ and $r$ are the slopes of angular frequency and redshift dependence of our power-law model (\ref{eq:IAmodel}). }
\label{fig:nz}
\end{figure}

\begin{figure}[h]
\centering
\includegraphics[width=9cm]{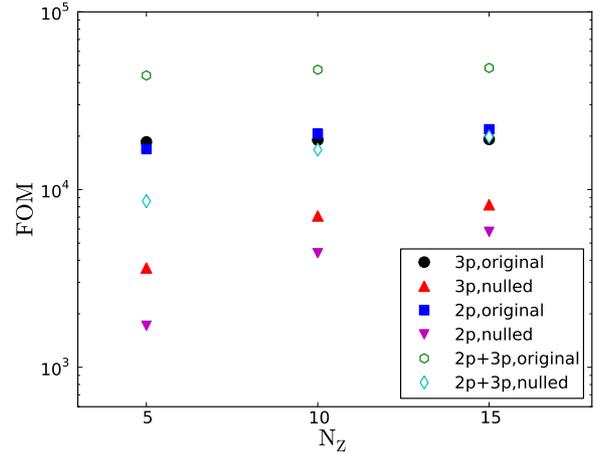}
\caption{ Figure of merit (FoM) as defined in (\ref{eq:fom}) in the $\Omega_{\rm m}$ - $\sigma_{\rm 8}$ plane as a function of number of redshift bins $N_z$. FoM from two-point (2p) measures, three-point measures (3p) and combined (2p+3p) are shown both before nulling (original) and after nulling (nulled). }
\label{fig:fom_nz}
\end{figure}

To re-assess this problem at the three-point level, we consider two different situations to address the requirements coming from control of the intrinsic-shear alignment and preservation of the information content separately. In both cases we split the redshift range between $z=0$ and $z=3$ into 5, 10, 15, and 20 (only in the first situation) redshift bins, with the redshift bins split in a way that there is an equal number of galaxies in each bin.

First we consider a single cosmological parameter, $\Omega_{\rm m}$, to be free and study the biases introduced by the GGI signal on $\Omega_{\rm m}$ both before and after nulling. We use only equilateral triangle configurations to reduce the amount of calculation. 
The results are shown in Fig.$\,$\ref{fig:nz}. Within the range of consideration, the ratio of the nulled and the original biases drops quickly with the increase of the number of redshift bins for all GGI models. For most of the models, 5 redshift bins seem to be not sufficient for the nulling technique to control the bias induced by GGI down to a percent level. Going from 5 redshift bins to 10 redshift bins is very rewarding in terms of bias reduction.
However, we note that a decrease $|\rm{b}_f/\rm{b}_i|$ doesn't neccessarily indicate a better performance of the nulling method, or generally speaking, of any method intended to control the intrinsic-shear alignment. One can see the reason for this by noticing that, it is the original unbinned GGI/GGG signal that is directly controlled by any of these methods. Between $|\rm{b}_f/\rm{b}_i|$ and the original unbinned GGI/GGG signal lies the binning process as well as the summation over angular frequency bins and redshift bins. Since the signs of the biases contributed by different angular frequencies and redshifts can be different, there can be bias cancellation during these processes. In another word, $|\rm{b}_f/\rm{b}_i|$ can depend on binning choices.

\begin{figure*}[t]
\begin{minipage}[c]{.75\textwidth}
\centering
\includegraphics[width=13cm]{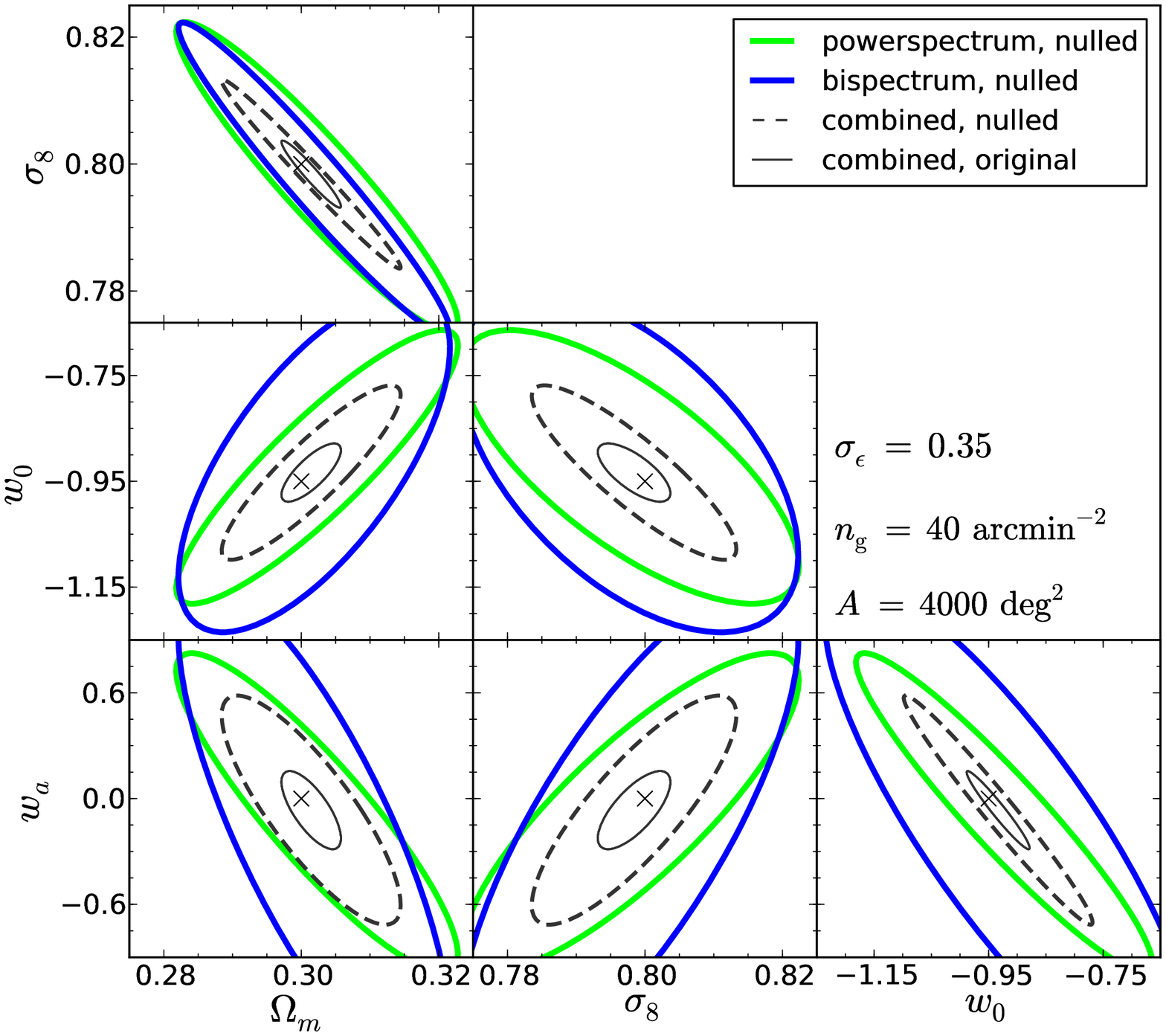}
\end{minipage}
\begin{minipage}[c]{.22\textwidth}
\caption{The thick green (gray) solid, thick blue (black) solid and thin black dashed ellipses indicate 1-sigma (68$\,\%$ CL) parameter constraints from the nulled power spectrum measures, bispectrum measures, and combined. Hidden parameters are marginalized over. The distance from the center of an ellipse to the black cross reflects the nulled bias on the corresponding parameter. The original biases from bispectrum measures can be seen in Fig.\ref{fig:para}. The thin black solid ellipses over-plotted on to the centers of the nulled combined constraint ellipses indicate the statistical power (68$\%$ CL) of combined constraints before nulling. Note the different ranges of parameters compared to Fig.\ref{fig:para}. 
}
\label{fig:combined}
\end{minipage}
\end{figure*}

We then vary two cosmological parameters ($\Omega_{\rm m}$ and $\sigma_{\rm 8}$) and investigate how the original and the nulled parameter constraints change with respect to the number of redshift bins available. For this case we use all triangle shapes to enable a comparison with results for two-point statistics.

Our result (Fig.$\,$\ref{fig:fom_nz}) shows that a further increase of the number of redshift bins beyond 10 is not very rewarding in terms of information preservation as characterized by the FoM, in either 2p, 3p, or 2p+3p cases. This suggests, when the possibility of more redshift bins exists, the choice of redshift bin number should be based mainly on the requirement of bias reduction level in case of negligible photometric errors. When there are non-negligible photometric errors, however, the information loss will probably be more severe, as found by \citet{joachimi09a} for the two-point case.

\subsection{The nulling technique as a conditioned compression of data}
\label{sec:compression}

The necessity of carrying out data compression in cosmology has long been recognized \citep[e.g.][]{tegmark97} and has been ever increasing due to the increasing size of the data sets. In cosmic shear studies the survey area of next generation multicolor imaging surveys will be an order of magnitude larger than the current ones. The study of three-point statistics also implies a huge increase in the amount of data directly entering the Fisher-matrix/ likelihood analysis, compared to the two-point case.    

\begin{table}[h]
\begin{minipage}[t]{\columnwidth}
\caption{Change of cosmic shear bispectrum statistical power after nulling (null) and linear data compression (compress).}  
\label{tab:FoM}
\begin{tabular}[t]{c|c|c|c|c|c}
  & i & null & null/i & compress  & compress/i \\
\hline\hline  
$\Omega_{\rm m}$-$\sigma_8$  & 21455  & 4609   &21.5$\,\%$  & 12242  &57.1$\,\%$\\
$\Omega_{\rm m}$-$w_0$       & 637    & 123    &19.3$\,\%$  & 428    &67.2$\,\%$\\
$\Omega_{\rm m}$-$w_a$       & 145    & 33     &23.0$\,\%$  & 110    &75.9$\,\%$\\
$\sigma_8$- $w_0$            & 434    & 87     &20.0$\,\%$  & 299    &68.9$\,\%$\\
$\sigma_8$- $w_a$            & 101    & 26     &25.4$\,\%$  & 72     &71.3$\,\%$\\
$w_0$- $w_a$                 & 11.4   & 2.3    &20.2$\,\%$  & 8.0    &70.2$\,\%$\\

\end{tabular}
\tablefoot{Presented are FoM on two-dimensional parameter planes between cosmological parameters $\Omega_{\rm m}$, $\sigma_8$, $w_0$ and $w_a$. The cosmological parameters $h$, $\Omega_{\rm b}$ and $n_{\rm s}$ are marginalized over. The second column is the FoM from the original bispectrum; the third and fifth columns are FoM from the nulled and the compressed measures, respectively; the fourth (sixth) column shows the percentage of the third (fifth) column compared to the first column, which reflects the relative information loss through the nulling (the unconditioned compression) procedure.}
\end{minipage}
\end{table}

The basic principle of data compression is to reduce the amount of data while preserving most of the information. This is already naturally encoded in the nulling technique. If one keeps only the first-order weights for nulling, as we do in this paper, the nulling procedure reduces the number of data entries in each angular frequency bin from the number of redshift bin $(i,j,k)$ combinations, to the number of $(i,j)$ combinations, which means roughly from ${N_z}^3$ to ${N_z}^2$. The nulling transformation is linear since the resulting nulled entry is a linear combination of $k$ original entries weighted by the nulling weight (\ref{eq:Yij}). In the sense that an ``optimum'' set of nulling weights is constructed, the nulling technique also intends to preserve as much information as possible. But there is yet another additional constraining condition in the nulling procedure: the nulling condition (\ref{eq:nc}), which largely confines the shape of the nulling weights by requiring the existence of at least one zero-crossing (see Fig.\,\ref{fig: poptimal}). In short, the nulling technique can be seen as a conditioned linear compression of data.

\begin{table*}
\centering
\caption{FoM before ($`i'$) and after ($`f'$) nulling and their ratio, using the cosmic shear power spectrum (2pt), bispectrum (3pt), and combined (2pt+3pt) analysis.} 
\label{tab:FOM23null}
\begin{tabular}[t]{c|c|c|c|c|c|c|c|c|c}
   & 2pt,\,i & 3pt,\,i & 2pt+3pt,\,i & 2pt,\,f & 2pt,\,i/f & 3pt,\,f & 3pt,\,i/f & 2pt+3pt,\,f & 2pt+3pt,\,i/f\\
\hline\hline  

$\Omega_{\rm m}$-$\sigma_8$ & 21774 & 21455& 86851 & 3297 & 15.1$\,\%$ & 4609  &21.5$\,\%$ & 18555 & 21.4$\,\%$\\
$\Omega_{\rm m}$-$w_0$      & 1590  & 637 & 3806  & 236  & 14.8$\,\%$ & 123  &19.3$\,\%$ & 600 & 15.8$\,\%$\\
$\Omega_{\rm m}$-$w_a$      & 517   & 145  & 872  & 69   & 13.3$\,\%$ & 33   &23.0$\,\%$ & 121 & 13.9$\,\%$\\
$\sigma_8$- $w_0$           & 864   & 434  & 3832  & 132  & 15.2$\,\%$ & 87  &20.0$\,\%$ & 488 & 17.2$\,\%$\\
$\sigma_8$- $w_a$           & 326   & 101  & 709   & 47   & 14.4$\,\%$ & 26   &25.4$\,\%$ & 107 & 15.1$\,\%$\\
$w_0$- $w_a$                & 45    & 11   & 184   & 7.4    & 16.4$\,\%$ & 2.3    &20.2$\,\%$ & 27  & 14.5$\,\%$\\
   
\end{tabular}
\renewcommand{\footnoterule}{}  
\end{table*}

It is then interesting to know how much of the information loss during the nulling process actually comes from the nulling condition, and how much just comes from the fact that a data compression process is naturally involved in nulling. To explore this, we perform an unconditioned linear data compression, by simply ignoring the nulling condition in the whole nulling procedure i.e. dropping the Lagrange multiplier term in (\ref{eq:maxi}), but otherwise keeping the simplifications inherent to the analytical approach. The results are shown in Fig.$\,$\ref{fig:para}. A summary of the FoM from the original and the nulled bispectrum measures as well as the compressed measures is shown in Table$\,$\ref{tab:FoM}. 

In contrast to nulling, an unconditioned linear compression does not eliminate the parameter bias, but increases or reduces some of them marginally. Regarding the parameter constraints, although the increase in the size of the ellipses is much less than in the case of nulling, around one third of the information in terms of FoM is lost through compression, which means that the amount of degradation in parameter constraints after compression is not negligible. This suggests that keeping only the first-order terms contributes to non-negligible information loss. To regain part of this information, one could add higher-order weights to the nulling procedure. But the difference between the nulled and the compressed FoM serves as an indication for the inevitable information loss through the nulling process, which is imposed by the nulling condition.

\subsection{Two-point and three-point constraints combined}
Besides constraining cosmological parameters using three-point cosmic shear alone, we investigate the combined constraints from both two-point and three-point cosmic shear measures. The performance of the nulling technique on cosmic shear power spectrum tomography alone and the resulting constraints on cosmological parameters were presented in JS08. For consistency, we use the same setting for the cosmic shear power spectrum as described for the bispectrum in Sect.\,\ref{sec:model}. In particular, we neglect photometric redshift errors, use only a limited range and number of $\ell$-bins, and adopt a power-law intrinsic-shear alignment model with a form described by (36) in JS08 and a slope of 0.4. We have confirmed the consistency between our power spectrum and bispectrum codes with those used in \citet{berge09}. Our power spectrum code agrees also with iCosmo \citep{refregier08}.

Figure \ref{fig:combined} shows the resulting constraint ellipses after nulling from the cosmic shear power spectrum analysis, the bispectrum analysis, and the two combined. To show how much information is lost during the nulling process, we overplot the original two- and three-point combined constraints on top of the nulled constraint ellipses in Fig.\,\ref{fig:combined}, but center them on the corresponding nulled constraints by subtracting the bias difference before and after nulling. The information content in terms of FoM for each parameter pair is presented in Table\,\ref{tab:FOM23null}.

One sees that the amount of information contained in bispectrum measures and power spectrum measures are indeed comparable. With bispectrum information added, typically three times better constraints in terms of FoM are achieved, both before and after nulling. This factor is smaller than the result in TJ04, although the same angular frequency range and the same set of 7 cosmological parameters are chosen for both studies. However a direct comparison is prohibited by different fiducial values adopted and different survey specification. 

Through the nulling procedure, around 15$\,\%$ of the original information in terms of FoM is preserved in the two-point case, and around 20$\,\%$ in the three-point case. It is a bit higher in the three-point case, in accordance to the fact that a roughly ${N_z}^3 \rightarrow {N_z}^2$ compression is involved in the three-point case and a ${N_z}^2 \rightarrow {N_z}^1$ one in the two-point case, while this fact is due to the summation over one redshift bin index during the nulling procedure (the same trend is evident in Fig.$\,$\ref{fig:fom_nz}). The information loss is considerable, but it is a price to pay for a model-independent method. As we have discussed in the previous subsection, the difference between the information loss through the nulling and the unconditioned compression procedures represents the inevitable loss of information through nulling. However, this difference is less than 50$\,\%$ in the considered three-point case. The other information loss is due to the simplifications we adopted in this study, including using only the first-order weights, and discarding the measures with two or three equal redshift bins. A further detailed consideration of these aspects can regain part of the lost information. Another simplification we have made in the three-point case is to use only triangles with three different angular frequencies. This reduces both the original and the nulled information contained in the three-point measures. However, this simplification can be easily removed with a careful distinction of all cases. 

Also notice that, the dependence of number of possible bispectrum modes, i.e. triangles, on the maximum angular frequency $\ell_{\rm max}$ is roughly $\ell_{\rm max}^3$, while that of power spectrum modes is roughly $\ell_{\rm max}^1$. For this study $\ell_{\rm max}=3000$ is chosen. If reliable information on smaller angular scales can be obtained, the three-point statistics will possibly give us more information than the two-point statistics.

\section{Conclusion}

In this study we developed a method to control the intrinsic-shear alignment in three-point cosmic shear statistics by generalizing the nulling technique. We showed that the generalization of the nulling technique to three-point statistics is quite natural, providing a model-independent method to reduce the intrinsic-shear alignment signals (GGI and GII) in comparison to the lensing GGG signal. 

To test the performance of the nulling technique, we assumed a fictitious survey with a setup typical of future multicolor imaging surveys, and applied the nulling technique to the modeled bispectra with intrinsic-shear alignment contamination. The lensing bispectra (GGG) was computed based on perturbation theory, while the GGI signal was modeled by a simple power-law toy model. We focused on the reduction of the GGI contaminant, since GII can be removed simply by not considering tomographic bispectra with two or three equal redshift bins.

The reduction of the intrinsic-shear alignment contamination at the three-point level by the nulling technique was demonstrated both in terms of the GGI/GGG ratio, and in terms of biases on cosmological parameters in the context of an extended Fisher matrix study. In terms of the GGI/GGG ratio, a factor of 10 suppression is achieved after nulling over all angular scales. Correspondingly, the biases on cosmological parameters are reduced to be less than or comparable to the original statistical errors. We studied the performance of the nulling technique when 5, 10, 15, or 20 redshift bins are available, and found that the performance on bias reduction, rather than how much information is preserved during the nulling procedure, depends more significantly on the number of redshift bins. In case one requires better control of intrinsic-shear alignment, more detailed redshift information allowing more redshift bins is the most direct way to go.
 
When dealing with real data, there is one further source of complication which we did not consider in this paper, that is the photometric redshift uncertainty. The photometric redshift uncertainty can be characterized by a redshift-dependent photometric redshift scatter and catastrophic outliers. \citet{joachimi09a} studied the influence of photometric redshift uncertainty on the performance of the nulling technique at the two-point level. They found that the photometric scatter  places strong bounds on the remaining power to constrain cosmological parameters after nulling. The existence of catastrophic outliers, on the other hand, can lead to an incomplete removal of the intrinsic (II, III) alignments as well as the intrinsic-shear alignments (GI, GII, GGI). However, methods to control the photometric redshift uncertainty have been proposed. For example, recent studies concerning the problem of catastrophic outliers point to the solutions of either limiting the lensing analysis to $z<2.5$ or by conducting an additional small-scale spectroscopic survey \citep{sun09, bernstein09, bordoloi09}.

As already demonstrated by JS08 in the two-point case, some information loss is inherent to the nulling procedure.
For the setup of this paper we found that, in terms of FoM about 20$\,\%$ of the original information is preserved through the nulling procedure in the three-point case, and 15$\,\%$ in the two-point case. We further studied the source of such information loss by comparing the nulling technique to an unconditioned linear compression of the data, since the nulling procedure can be seen as a linear compression of data under the constraint of the nulling condition (\ref{eq:nc}). We found that around one third of the original information is lost through an unconditioned compression of the data, suggesting that this situation can be improved by considering higher-order terms in the nulling and compression processes. 

Results on parameter constraints from the two- and three-point cosmic shear statistics combined are also presented. The amount of nulled information contained in bispectrum measures and power spectrum measures are comparable. With bispectrum information added, typically three-times better constraints are achieved both before and after nulling, in terms of FoM.  

Again, due to the large amount of information existing in the three-point cosmic shear field, one would certainly like to exploit it in the future. The nulling method we developed in this work solves a potentially severe problem hampering the use of three-point information, namely the intrinsic-shear alignment systematic. Our method works at the cost of a large information loss, which can hopefully be avoided by a future method of removing the intrinsic-shear alignment contaminants. But as the only completely model-independent method so far, the nulling technique can serve as a working method now and can provide a valuable cross check even with the availability of better methods.

\begin{acknowledgements}
We thank Joel Berg{\'e} and Masahiro Takada for the help in comparing power spectrum and bispectrum codes. We are also grateful to the anonymous referee for the helpful comments and suggestions. BJ acknowledges support by the Deutsche Telekom Stiftung and the Bonn-Cologne Graduate School of Physics and Astronomy. This work was supported by the RTN-Network `DUEL' of the European Commission, and the Deutsche Forschungsgemeinschaft under the Transregional Research Center TR33 `The Dark Universe'.

\end{acknowledgements}

\bibliographystyle{aa}
\bibliography{bibliography}

\begin{appendix}

\section{Counting of triangles}
\label{app:1}
\begin{figure}[h]
\centering
\includegraphics[width=9cm]{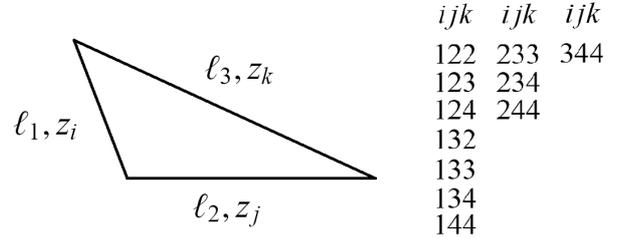}
\caption{List of possible triangles (redshift bin combinations) with condition $z_i<z_j$ and $z_i<z_k$ when 4 redshift bins are available. An angular frequency combination satisfying $\ell_1 \le \ell_2 \le \ell_3$ is chosen. Note that the redshift indices and the angular frequencies are linked in pairs due to the definition of the tomographic bispectrum (\ref{eq:fbisp}). In this paper a default of 10 redshift bins is assumed. 
}
\label{fig:count}
\end{figure}

A triangle is specified by six indices, i.e. three redshift bin indices \{$i, j, k$\} and three angular frequency bin indices $\ell_1, \ell_2, \ell_3$. To ensure that we count each triangle configuration only once, we set the condition that $\ell_1 \le \ell_2 \le \ell_3$. Moreover, we would like the first index among \{$i, j, k$\} in (\ref{eq:F_i}) to have the lowest redshift, i.e. $z_i<z_j$ and $z_i<z_k$, for the convenience of performing the nulling technique. The possible \{$i, j, k$\} combinations under these constraints in the case of $N_z=4$ are listed in Fig.$\,$\ref{fig:count}.

However, setting both conditions is problematic. Inspecting the definition of the tomographic bispectrum (\ref{eq:fbisp}), one sees that the redshift indices and the angular frequencies are linked in pairs, e.g. convergence $\kappa$ in redshift bin $i$ has angular frequency ${\ell}_1$, which is not desirable since the smallest angular scale does not necessarily correspond to the lowest redshift. To solve this problem, we perform nulling three times for each general angular frequency combination with $\ell_1<\ell_2<\ell_3$, swapping the redshift-angular scale correspondence in-between, thus allowing each redshift to be able to correspond to any angular frequency. 

Note that the situation complicates a bit when two of the angular frequencies are equal, since then the swapping may lead to exactly the same configuration. To avoid this, we will restrict ourselves to three different angular frequencies. This can exclude a high percentage of possible configurations. In our case, i.e. 20 logarithmically spaced bins between $\ell_{\textrm{min}}=50$ and $\ell_{\textrm{max}}=3000$, 37$\,\%$ of the angular frequency combinations which can form a triangle have been excluded. However, this is only a technical complication which can be solved with a careful distinction of all cases. Since this paper is intended to be a proof of applicability of the nulling technique to three-point statistics, we defer the intricacies of accounting for all triangle configurations to future work.

\end{appendix}

\end{document}